\documentclass{article}
\usepackage{iclr2026_preprint,times}

\usepackage{amsmath,amsfonts,bm}

\def\eqref#1{equation~\ref{#1}}

\def\1{\bm{1}}

\DeclareMathAlphabet{\mathsfit}{\encodingdefault}{\sfdefault}{m}{sl}
\SetMathAlphabet{\mathsfit}{bold}{\encodingdefault}{\sfdefault}{bx}{n}

\usepackage{hyperref}
\usepackage{url}
\usepackage{booktabs}
\usepackage{array}
\usepackage{graphicx}
\usepackage{tikz}
\usetikzlibrary{arrows.meta,positioning,calc,fit}
\usepackage{amsmath,amssymb}
\usepackage{cleveref}
\hypersetup{pdfauthor={Jin Han, Mingfeng Xin, Feng Liang, Jingtao Yu}, pdftitle={NIOCIS: Typed-Decision Next-Hop Routing with Node-Conditioned Option Spaces for Tool-Use Agents}}

\newcommand{\method}{NIOCIS\xspace}
\newcommand{\graphname}{loop-collapsed action graph\xspace}
\newcommand{\DAG}{DAG\xspace}
\newcommand{\router}{typed-decision router\xspace}
\newcommand{\pnum}[1]{\textbf{P#1}}
\usepackage{xspace}
\DeclareMathSizes{9}{9}{7}{5}

\title{NIOCIS: Typed-Decision Next-Hop Routing\\ with Node-Conditioned Option Spaces\\ for Tool-Use Agents}

\author{Jin Han \quad Mingfeng Xin\bf\\
\bf Feng Liang \quad Jingtao Yu\\
Independent Researcher}

\iclrfinalcopy
\begin{document}
\lhead{Preprint}

\maketitle
\begin{abstract}
Tool-use agents pay per decision: at each hop a language model re-reads the state
and thinks for seconds, though its actions are finite and its decisions recur. We
present \method{}, a routing layer that folds execution logs into a \graphname{}---forward
action transitions plus explicit recovery back-edges---and answers one typed
question: which declared option at a state, plus \texttt{stop}, does the fleet take
next? A small model reads a calibrated distribution over them in one pass, generating
no text. On 1{,}615 production security-operations episodes we measure what the layer
replaces, returns and is bounded by. It \emph{replaces} the per-decision deliberation:
the typed encoder matches the logged next hop on 32.2\% of the 3{,}875 held-out
decisions (34.9\% within the declared options) in 10\,ms of encoder time (19.5\,ms for the
typed-choice form), against 2.2\,s per LLM hop; two feature readouts on the same decisions
reach 32.8\% and 32.3\%, an ordering the paired test does not resolve, so the typed layer is
chosen for cost rather than for a separated accuracy lead.
Tuning moves a language model: jpt-4b reaches 9.9\% zero-shot and 31.9\% fine-tuned, a
gap an exact paired test does not resolve ($p=0.68$), at 244\,ms per decision. It
\emph{returns} little: the realised hop falls outside the declared options for 7.9\%
of decisions and outside the model's top twenty for 9.3\%, while the most confident
15\% holds 91.9\% accuracy, a band three different confidence rules trace alike. And
it is \emph{bounded}: the state text binds accuracy, not the candidate count, the
deliberation budget or the parameter count; node discovery saturates at the rate a
zero-parameter occupancy null predicts, and on five public process logs a
variable-order Markov table collects most of each margin over the majority baseline; a
cross-trace leave-one-out test puts the released state's ceiling at 25.6--44.9\% as
the state key grows more specific, repeated states disagreeing more often than they
agree; and the option-space ablation attributes the gain to per-node candidate
restriction (+3.5 points over a flat catalog) rather than graph traversal (zero within
its interval). The claim is narrow: containment and a small high-precision band at
millisecond cost; a gated self-evolution loop lifts every arm above its anchor yet ends
below full supervision; a confidence-gated cascade \emph{loses} agreement when the
escalation arm is the zero-shot stand-in rather than the tuned model; and
agreement past a third of the fleet needs a state carrying the plan and the
observation, which the released records do not.
\end{abstract}

\section{Introduction}
\label{sec:intro}

LLM agents that call tools decide by inference. At every step of every task
the model re-derives what it already decided last week for the same tool in
the same situation: whether to search before reading, whether to write
directly or route through approval, whether to retry or roll back. The
latency of such a judgment is counted in seconds and its cost in thinking
tokens~\citep{cheng2026thisthat,li2026jevedge}, even though the decision's
\emph{support} is small---a closed set of declared next actions,
median 46 per state here---and the
deployment's accumulated logs already contain the answer, thousands of times
over.

Three observations motivate \method{}. Execution logs converge.
An internal deployment's tools and action semantics are finite; as episodes
accumulate, the action vocabulary saturates while
the logs keep growing. Process mining has exploited this regularity for business
event logs for two decades~\citep{vanderaalst2016processmining,leemans2013inductive},
and recent results compress agent-trace corpora into finite-state
machines of 7--43 states~\citep{cho2026automata}; the complexity-vs-log-size reading we cite tracks only the state-count curve (\S\ref{sec:related}). Also,
typed decisions can be cheap. A decision model that answers exactly one of
its caller's declared options---reading a calibrated distribution from a
single forward pass, with no generated text---decides in 30--65\,ms~\citep{cheng2026thisthat,ren2026jelite}
at accuracy within a few points of LLM judges~\citep{ren2026jelite},
and works as a first-tier agent decision layer that escalates on low
confidence~\citep{wu2026reflex,li2026jevjudge}. And logs can train their
own router: routing decisions, executions and outcomes form a data
flywheel~\citep{liu2026agenticrouting,neo2026neohorse}, provided the update
loop is gated against collapse~\citep{nguyen2026recursiveheldout,shang2026skilaa}.

\begin{figure}[t]
\centering
\begin{tikzpicture}[
  font=\small,
  box/.style={draw, rounded corners=1pt, minimum height=9.6mm, minimum width=25mm, align=center, fill=gray!6},
  arr/.style={-{Stealth[length=2mm]}, thick}]
  \node[box] (log) {episode logs\\(retries, back-returns,\\rollbacks)};
  \node[box, right=5mm of log] (ind) {loop folding +\\graph induction};
  \node[box, right=5mm of ind] (gr) {\graphname\\$E^{+}$ forward + $E^{-}$ recovery};
  \node[box, below=9mm of ind] (rt) {\router{} $\pi_\theta$\\typed readout over $A(v)$};
  \node[box, below=6mm of rt] (ex) {executed walk\\budget $B$, revisit cap $R$};
  \node[box, right=16mm of rt] (llm) {LLM escalation\\(low confidence)};
  \node[box, right=5mm of ex] (up) {accumulate + rollback\\gate};
  \draw[arr] (log) -- (ind);
  \draw[arr] (ind) -- (gr);
  \draw[arr] (gr.south) -- ++(0,-4.5mm) -| (rt.north);
  \draw[arr] (rt) -- node[pos=0.58, above=2mm, font=\scriptsize] {$\max\pi{<}\tau$} (llm);
  \draw[arr] (rt) -- (ex);
  \draw[arr] (ex) -- (up);
  \draw[arr] (up.north) -- ++(0,3.5mm) -| (rt.south east)
        node[pos=0.9, right=1.5mm, font=\scriptsize] {retrain $\theta$};
  \draw[arr, dashed] (ex.west) -- ++(-6mm, 0) |- ([yshift=-3mm]log.south) -- (log.south)
        node[pos=0.45, left=1.5mm, font=\scriptsize] {new episodes};
\end{tikzpicture}
\caption{\method{}: logs are folded into a \graphname{}; a typed-decision
router answers ``which edge in $A(v)$ does the fleet take next?'' per
state, escalating only
uncertain cases to the LLM; executed walks flow back as gated updates.
The step budget $B$ and per-node revisit cap $R$ on the execution box are
full-design parameters: the corpus study of \S\ref{sec:setup} replays
logged episodes teacher-forced and exercises neither bound.}
\label{fig:system}
\end{figure}
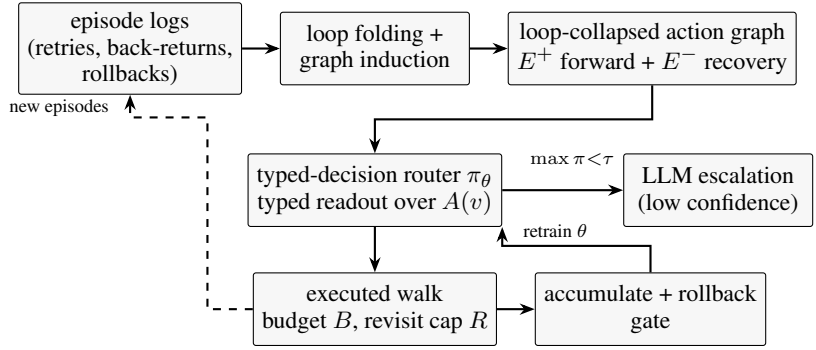

\method{} composes these into one loop (\Cref{fig:system}): (i)~we \emph{induce}
a \graphname{} from episode logs---a directed graph of forward action
transitions
(self-loops aside) with recovery back-edges carrying rollbacks, backtracks
and task redirects;
(ii)~we \emph{route} with the \router{} whose option set is exactly
the current node's outgoing edges (plus \texttt{stop}), whose output is a calibrated distribution
the caller can threshold, and whose uncertain cases escalate to the LLM;
(iii)~we \emph{self-evolve}: each real episode becomes training data, updates
pass a rollback gate and survive only with evidence, and any update that
regresses beyond the gate tolerance ($\delta$) is rolled back. In the full design, executed episodes are budget-bounded (a step
budget and a per-node revisit cap; \S\ref{sec:problem}). Our contributions are:
\begin{itemize}
  \item \pnum{1}~\textbf{The loop-collapsed action graph.}
  We formalize induction from episode logs with explicit folding rules for
  loops and back-edges, and quantify the growth of the induced action graph (E1):
  transition discovery shows diminishing returns and loop-event rates reach
  a stable regime, while node-vocabulary convergence is \emph{not} offered as an
  independent finding---a zero-parameter occupancy null already predicts it
  (\S\ref{sec:results}).
  \item \pnum{2}~\textbf{Typed-decision path routing.} We define next-hop
  selection as a typed decision over a node-conditioned declared option space
(the current node's outgoing edges, plus \texttt{stop}) and implement it
with one-pass feature-MLP readouts that reach 32.8\% agreement with the
deployed fleet's own choices (35.6\% within the declared space; the
hidden-state probe variant gives up two points of masked accuracy and
decodes from its 10\,ms encoder pass, the typed family's full predict
running 19.5\,ms p50), together with an accept/escalate gate analyzed with
risk--coverage curves (E2); a transition-table baseline over all nine
decision streams (App.~B) reaches 31.1\% on SOC. We also measure what the
restriction is worth (E9): the node-conditioned space beats a flat
155-class catalog by 3.5 points and a size-matched global-frequency space by
1.4 points, but is statistically indistinguishable from a node-conditioned
space built without graph traversal, so we claim the per-node restriction, not
graph traversal. We therefore claim the formalization and the
  measurement rather than the traversal: the contribution is to define the next-hop choice
  as a typed decision over a node-conditioned declared option space, and to bound what that
  construction can reach on production logs, which \Cref{sec:replace} does quantitatively.
  \item \pnum{3}~\textbf{A gated self-evolution loop, with route replay as a
  measurement.} We replay held-out episodes against a re-prompted MIMO Pro V2.6 stand-in (one
  reasoning call per hop for the fleet's LLM component: no per-stage orchestration, no tool results;
  \S\ref{sec:setup}); the route view renders the per-decision quantities---strict
  survival, plus a wrong-hop cost rescaling of the mismatch count---so we report it
  as a measurement on that graph (\S\ref{sec:results}), and add the one route
  quantity that is not such a rescaling---the first-divergence hop index and type
  (median hop 2, most first divergences on escalated hops)---reading the replay as placing the
  cheap readout in the agreement/cost corner and the accept/escalate
hybrid between its parents (the readout and the opponent; E3); and
we exercise a self-evolution loop
whose rollback gate scores every candidate while the loop absorbs a
12\%-of-one-batch label-corruption stress, the gate refusing 13 of the 40 candidate updates, in 6 of the 10 seeded chains (E4).
\end{itemize}

The core experiments (E1--E4) run on real execution logs from a production
security-operations agent fleet (\S\ref{sec:setup} states provenance);
E5 replicates the constructions on public event logs and open tool-use
trajectories, and any Langfuse-style observations export folds into the
same episode schema.

\section{Related Work}
\label{sec:related}

\paragraph{Workflow and skill induction from agent logs.}
Agent Workflow Memory~\citep{wang2024agentworkflowmemory} induces reusable
routines from trajectories and feeds them back as context; task-model
induction~\citep{jiang2026taskmodel} builds hierarchical goal/procedure models
from computer-use traces; automata induction compresses trace corpora into
compact finite-state machines for next-step prediction and failure
detection~\citep{cho2026automata}; the literature-search agent
\emph{PaperPilot} induces editable search-operator directed acyclic
graphs (DAGs)~\citep{li2026paperpilot}. Skill-graph systems organize experience for
\emph{retrieval}: HiSkill's hierarchical skill
graphs~\citep{hao2026hiskill}, SE-GoS's self-evolving
graph-of-skills~\citep{fu2026segos}, CaSKG's counterfactual-causal skill
graphs~\citep{li2026caskg}, GraphSkillAA's validated skill-graph
updates~\citep{shang2026skilaa}, GraphSkillEvo's evolutionary graph
skills~\citep{sun2026graphskillevo}, and skill distillation from browsing
traces~\citep{yang2026skillbc}. M2-PALE applies process mining to agent traces
for \emph{explanation}~\citep{qian2026m2pale}; AgentHER recycles trajectories
into training data~\citep{ding2026agenther}. Every workflow and skill system surveyed here uses the induced structure for retrieval, prompting, explanation or training (the
automata-induction uses of the same structure are treated next). Automata-based next-step prediction and failure
detection over induced
machines~\citep{cho2026automata} are the closest runtime uses: both
score a fixed label vocabulary from trace prefixes (held-out scoring
offline against the induced machine; its failure ranker also runs as an
online monitor over partial traces) rather than binding the
induced graph's edges as declared options. Interleaved reason-and-act control
loops keep an LLM in the loop at every step~\citep{yao2023react}, self-reflective
retry feeds verbal feedback back into the next attempt~\citep{shinn2023reflexion},
and open-ended agents grow executable skill libraries as text~\citep{wang2023voyager};
we instead bind the induced graph as the decision space of a single-pass readout,
with no per-step LLM call. None of the
surveyed systems makes the induced graph the
\emph{runtime decision space} of a millisecond-scale model, and none reports
route-level divergence accounting (episodic survival and wrong-hop cost)
for routes over an induced graph on production agent logs (the
accounting scores realized routes, not a model's conformance
fitness~\citealp{vanderaalst2016processmining}; alignment-based conformance
checking instead replays each trace against a discovered model and scores the
cheapest replay~\citep{adriansyah2011conformance,carmona2018conformance}). Read
as a discovered model the induced graph is a permissive one---on held-out
traces, token-level replay fitness $0.91$, trace-level $0.59$, escaping-edge
precision $0.02$ (each state enables its whole declared option set),
edge-level generalization $0.27$ (App.~B)---which is why our comparison is
against option-space constructions, leaving process-discovery models aside.
\citet{paul2026agentrouter} study cost-optimal multi-step
model/tool routing (a different decision axis).

\paragraph{Process mining.}
Discovery of process models from event logs is a mature
field~\citep{vanderaalst2016processmining}; constructive miners capture loop
and concurrency structure~\citep{leemans2013inductive,kourani2025partialorder},
and monotonicity characterizations control candidate-place
growth~\citep{tacke2026speccpp}. The complexity-vs-log-size reading we cite
reports stable small automata (7--43 states on public corpora,
\citealp{cho2026automata}); it carries no
loop-event or transition-pair axis. E1 adds those axes on a production
security-operations-center (SOC) corpus, with subsample-median bands, and measures the loop structure
directly. A parallel line, predictive process
monitoring, answers the next-activity question by scoring a fixed activity
vocabulary from prefixes~\citep{tax2017predictive}; its standard n-gram
realization is the backoff-Markov baseline of E5 (re-gated here onto the
shared declared option sets).

\paragraph{Typed-decision models and LLM routing.}
Typed-decision models answer exactly one of their caller's declared options
from a single forward pass, with calibrated probabilities and no generated
text: this-that-model-1.0 reads answers from hidden state at
30.9\,ms~\citep{cheng2026thisthat}; the one-pass judge we label \emph{JevLite}
(small one-pass model of~\citet{ren2026jelite}) reads
temperature-scaled answer-label logits at 64.5\,ms per decision on one consumer GPU. As judges they accept
when confident and escalate when unsure~\citep{li2026jevjudge}; as agent
control layers they cut strong-model calls 72.7\% at 95\% task success
against a strong-only baseline of 88\% on the same frozen
benchmark~\citep{wu2026reflex}; at the edge they cut decision latency and API
fees~\citep{li2026jevedge}, and with an LLM planner they close fast control
loops~\citep{ma2026jevstar}. Unlike these flat typed-control layers' static
action-list option sets, \method{}'s option set is the outgoing
edge set of an \emph{induced} action graph; \method{} also performs
  route-level divergence accounting over that graph, quantifies its
  node-vocabulary convergence, and exercises a gated self-evolution loop,
  none of
which flat typed-control layers do. Guided or
grammar-constrained decoding also restricts an output space, but it filters
a generator step by step~\citep{geng2023grammar}; the typed readout instead
emits one distribution over the declared options in a single pass.
Constrained decision heads are sensitive
to option labeling~\citep{sun2026typesafe}, which motivates our edge
identifiers. Model/tool routers choose \emph{which solver}
runs~\citep{chen2023frugalgpt,ong2024routellm,paul2026agentrouter,liu2026agenticrouting,bala2026multiagent}
(or generate retrieval configurations~\citealp{pan2026brane}), not which
graph edge advances a task; preference-trained small routers have cut
cost by over $2\times$ at matched quality in some evaluated
settings~\citep{ong2024routellm}; typed artifacts share
tool-routing knowledge across agents~\citep{chakraborty2026typed}; and router infrastructure now spans development, evaluation and
deployment, where lightweight routers prove more competitive under
tight-cost constraints~\citep{feng2026llmrouter}, though routing
accuracy plateaus below the oracle~\citep{lu2026routingplateau}. Amortized
planning replaces search with direct action inference in world
models~\citep{nguyen2026latentgeometry}; we amortize \emph{workflow-path}
decisions into a declared-option readout over an induced action graph.

\paragraph{Self-evolving agents.}
Routing decisions and executions can train their own
successors~\citep{liu2026agenticrouting,neo2026neohorse}
as long as the loop is gated: unguarded self-evolution can collapse or
suffer security drift~\citep{nguyen2026recursiveheldout,li2026auditingfin},
while validated updates with rollback keep graphs
healthy~\citep{shang2026skilaa} and recursive-in-recursive designs keep
harness and weights co-evolving~\citep{xue2026sciencebuddy}. Our E4 adds a
minimal complete instance---accumulate, gate on held-out performance, roll
back on regression---and deploys it under label corruption that the
gated loop absorbs on average (the refusal path refuses 13 of the 40 candidate updates, in 6 of the 10 seeded chains).

\section{Problem Formulation}
\label{sec:problem}

\paragraph{Episode logs.}
A log is a set of episodes. An episode is a sequence of steps
$e = \langle s_1, \dots, s_T \rangle$ where each step $s_t$ records an action
label $L(s_t) \in \mathcal{V}$ (an action semantic such as
\texttt{db.write}), a step kind (advance, retry or back: a back step revisits an earlier
label, opening a return edge in $E^{-}$) and a context
vector $\mathbf{c} \in [0,1]^d$ (full-design schema: urgency,
dependencies, authorization,
complexity, \dots). Episodes carry a start, a goal---in the full design
a node
$g \in \mathcal{G} \subseteq \mathcal{V}$---and an outcome.

\begin{figure}[t]
\centering
\includegraphics[width=0.98\linewidth]{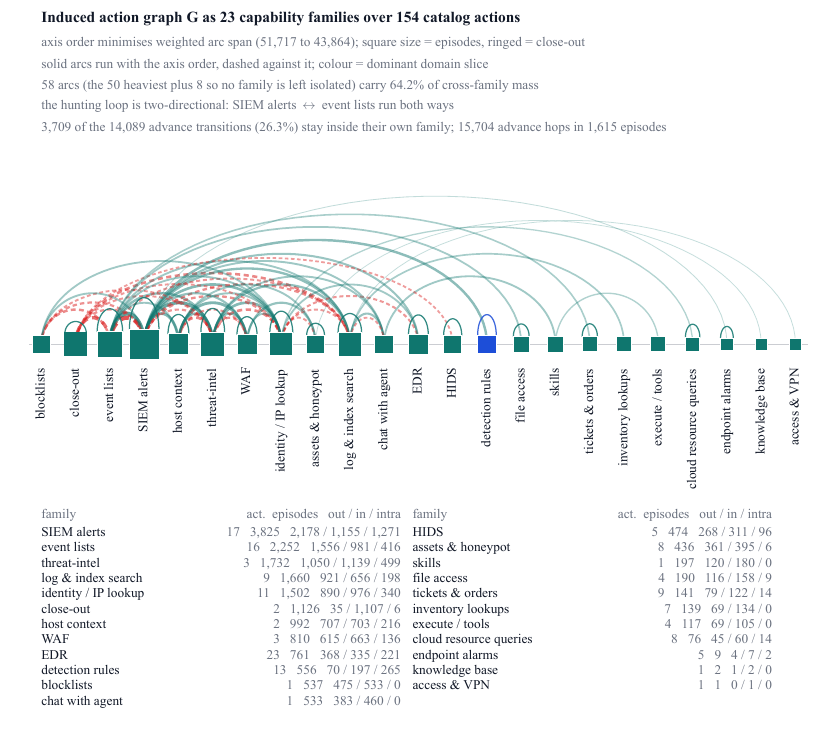}
\caption{\textbf{The induced action graph $G$ at the level on which every node can be named: a
capability-family structure over the released corpus.}
The 154 catalog actions are grouped into 23 capability families by an ordered
capability-keyword rule (a \emph{display quotient} defined in the released analysis code, not a
construct of the method), so the drawing covers all 154 stations and all 15{,}704 advance
hops with no anonymous station. The families are placed on one axis in the order that minimises
weighted arc span over the family pairs (51{,}717$\to$43{,}864); solid arcs run with that
order, warm dashed arcs against it, thickness proportional to transferred hops, and a hook on a node
carries the hops that never leave its family. Node size is episode count and colour the dominant
domain slice --- a deliberately honest non-finding: 22 of the 23 families are
detection-dominant, so colour separates one node. Two properties of $G$ that a station-level view
hides become visible content. First, the hunting loop is \emph{two-directional}
($\text{SIEM alerts}\leftrightarrow\text{event lists}$ 550/465,
$\text{identity}\leftrightarrow\text{threat-intel}$ 178/170, $\text{host}\leftrightarrow
\text{threat-intel}$ 171/142). Second, 3{,}709 of the 14{,}089 advance transitions
(26.3\%; the 15{,}704 advance hops of \S\ref{sec:problem} less one opening step per
episode---1{,}615 episodes) never leave their own family, 1{,}271 of them in the SIEM family
alone. The 58 drawn arcs are the 50 heaviest family-to-family exchanges plus one arc for
each family the budget would otherwise leave isolated, and they carry 64.2\% of cross-family
mass; the remaining share is counted, never implied absent, in the table, which lists every family
with its station count, episodes and out/in/intra-family mass. \textbf{close-out} (ringed) is the
graph's only terminal group: every hunting line is resolved there.}
\label{fig:graphshape}
\end{figure}

\paragraph{Loop-collapsed action graph.}
Straight traces understate structure; agents loop. Folding an episode yields
a skeleton---retry runs folded as temporally extended steps in the sense
of~\citet{sutton1999mdps}---to which the
step's label $L(s_t)$ is appended except in two step kinds: (a)~an
immediate retry of the last label kept in the skeleton (the retry is
dropped), and (b)~a revisit of an earlier-visited label (the step is
appended as usual and a new kept frame---a skeleton entry---opens at the
revisit). Both kinds are recorded as
\emph{loop events}
(\emph{retry} for (a), \emph{back} for (b)); a retry---which includes
every continuation of a same-label run---is the only kind left out of
the skeleton. Folding the 22{,}192 logged tool calls thus drops exactly
the 4{,}124 retries and keeps 15{,}704 forward advances plus 2{,}364
recovery returns (18{,}068 decision states); E1's event rates reflect
this disposition. Observed kept-frame transition pairs
($E^{+} \cup E^{-}$) over all episodes define
the transition volume reported in E1; first-visit forward transfers give $E^{+}$
and executed returns to earlier labels give \emph{recovery back-edges} $E^{-}$ (this corpus-census set is computed over all folds and
persisted with the corpus for characterization only; the operational
option graph is induced from the training fold, \S\ref{sec:method}), while same-label repeats surface only as \emph{retry}
events, as described above. The two roles a
pair can play---forward transfer and recovery return---overlap on real
traces: of the 2{,}536 distinct observed pairs, 2{,}315 appear as
first-visit forward transfers and 702 as recovery returns (481 pairs
play both roles---forward in one context, return in another---while 221
pairs are exclusively returns; App.~A gives the fold breakdown).
The \graphname{} is $G = (V, E^{+} \cup E^{-})$.
The observed forward set is not a \DAG{}: 1{,}272 of its 2{,}315 pairs
fall into 636 mutual couples ($u{\to}v$ and $v{\to}u$ between nodes $u$ and $v$,
both arcs observed as forward transfers), so forward transfers alone carry directed
2-cycles. What planning needs still holds: all edge weights are strictly
positive, so an optimal walk to a reachable goal can always be chosen
simple (no repeated node), even though execution may loop. Complexity is
measurable two ways---folded (semantic nodes and transitions) and
occurrence-tagged (per-kind step occurrences)---and \emph{loop mass} is
an occurrence-tagged quantity: the loop-event share of executed
steps, $100(\mathrm{retry}+\mathrm{back})/\mathrm{steps}$ (29.2\% of
the SOC corpus's steps). \Cref{fig:graphshape} draws the same graph one level up: the catalog
actions grouped into capability families, with the family-to-family traffic that a station-level
view hides.

\paragraph{Next-hop decision problem and route-level criterion.}
Let $\pi(a \mid v, g, \mathbf{c})$ be a router at the current node $v$ over the declared option set
$A(v) = \mathrm{succ}(v) \cup \{\texttt{stop}\}$ (options identified with
their edges). The reference next hop is the deployed fleet's
logged choice at the same realized state; the full design substitutes the
next hop of a context-weighted shortest-path walk to the goal $g$ for that
logged choice (a
guard fact $(a, \eta)$ pairs an action $a$ with a context precondition over
slot $\eta$, say an authorization level; a failed fact multiplies the
cost $w_e$ of the edge labeled $a$ by a penalty factor $\gamma>1$, so
a walk whose context is at a low authorization level is routed through
an approval detour). Per step we report top-1 agreement with the
reference. At the route level we report \emph{strict survival}---an episode's
fidelity to this one realized reference trace, where survival requires every one
of its decisions to match (a trajectory-fidelity reading, separate from any
fleet-level safety or availability property)---and the \emph{wrong-hop gap}, the
accumulated cost of diverging hops over the whole teacher-forced episode
under per-hop conventions $c_1 = 1$ and $c_3 = 3$; clipping that
accounting at the first divergence would understate the cost of long
episodes, so the first-divergence (strict-survival) view and the
unclipped wrong-hop gap are reported side by side. Under this
teacher-forced accounting the gap is a fixed rescaling of the per-decision
mismatch count---$(\#\text{decisions}/\#\text{episodes})\,(1-\text{agreement})$---so
its extra information over per-decision agreement is the episode-length
weighting; strict survival is the genuinely route-level measure.

\paragraph{Typed decisions and gated escalation.}
The full typed design's output is a distribution over $A(v)$ and nothing else: options
are structurally identified (stable edge identifiers, never surface names,
in the full design),
the answer is read from one forward
pass (as in \citealp{cheng2026thisthat,ren2026jelite}), and malformed outputs are
unrepresentable (the corpus readout emits over the global vocab then
masks to $A(v)$, cf.\ A3). A confidence gate accepts at threshold $\tau$ when
$\max_a \pi(a \mid \cdot) \ge \tau$ and otherwise escalates to the LLM,
mirroring accept/judge cascades~\citep{li2026jevjudge,wu2026reflex}. Since a
valid output is not a guarantee of progress, the full design runs execution under a step budget $B$ and a per-node revisit cap $R$, terminating the episode (failure) on either violation. The corpus study of \S\ref{sec:setup} instead replays logged episodes teacher-forced and so exercises neither bound.

\section{The \method{} Approach}
\label{sec:method}

\subsection{Graph induction and loop folding}
Folding follows the two rules of \S\ref{sec:problem}; the induction pass
book-keeps skeleton transitions and loop events incrementally per episode.
Recovery back-edges discovered during execution---rollbacks, backtracks
and task redirects---are recorded alongside the corpus; the
\S\ref{sec:setup} census over them ($E^{+}$, $E^{-}$) is computed over
all folds and is purely descriptive. The operational option graph, by
contrast, is induced from the training fold alone and is one shared
object: routers and rollouts see the same train-induced option graph at
training and at evaluation time, with no held-out edge ever entering the
runtime candidate set (see Option supply below). We deliberately do not
merge near-synonymous actions---semantic action labels are the unit of
decision, matching how typed decisions declare their
options~\citep{cheng2026thisthat}. Because option surfaces are load-bearing
for constrained heads~\citep{sun2026typesafe}, edge options are rendered over
stable internal identifiers rather than surface names, in the full design. (\emph{Option supply.} In the corpus
realization the declared option set at $v$ is the train-split fold's
kept-frame transitions out of $v$, first-visit advances and recovery returns alike, plus \texttt{stop}---graph induction and
routing share one graph, so the option space carries the induction's
misses, 7.9\% of test targets falling outside it---\S\ref{sec:setup};
\mbox{\Cref{fig:convergence}a} reports the observed-transition volume this
induction processes.)

\subsection{Typed-decision readout router}
We call the model the \router{}; its scoring head is the \emph{readout}.
The router answers one question per state: \emph{given the current node
$v$, goal $g$ and context $\mathbf{c}$, which edge in $A(v)$ leads to task
completion?} ($A(v)$ includes \texttt{stop} as a non-edge option.) In the corpus realization $g$ is the episode's
agent-stage identity---constant within an episode, hardwired to 0 on the
E5 streams---and $\mathbf{c}$ is structural (prefix shape, visit
counters, per-node neighbor statistics); the guard-aware $\mathbf{c}$ of
the full design feeds only the substituted context-weighted walk.
Options are the declared outgoing edges, bound to stable
single-token identifiers: target-action classes over the corpus's
154-name tool catalog (plus stop) in the class-space readout, or
edge identifiers of the induced graph in the edge-ID
readout (\S\ref{sec:results} compares both); in the hidden-state arms the answer distribution is
read from the hidden
state at a designated position (as in \citealp{cheng2026thisthat,ren2026jelite})---equivalently, a softmax over the declared
label logits (temperature-scaled only in our zero-shot
arm)---while the feature arms parameterize the same
softmax over observable features (the edge-ID readout space applies to both
backbones); nothing is generated.
Training minimizes proper-scoring losses over the full vocabulary
softmax on (state, goal, context, next-hop) tuples copied from held-in
logged transitions---behavioral cloning~\citep{pomerleau1991navigation} of the deployed fleet
(out-of-space targets stay in the objective; option masking enters only
at the masked-renormalized (masked-renorm) gate at decode); the full design
replaces the logged next hop with planner-oracle labels---the hop of a
context-weighted shortest-path walk to the goal---and uses the
walk-membership and episode-success questions as the oracle's post-hoc
reward signal. Calibration is reported as expected
calibration error (ECE-15, the 15-bin convention; temperature-fitted~\citealp{guo2017calibration} for the zero-shot arm,
raw masked-renorm confidence for all trained arms, over the probe and
feature backbones and their class-space and edge-ID variants).
Plain
cross-entropy (CE) and the CE-plus-Brier (CE$+$Brier)
objective~\citep{ren2026jelite} are compared head-to-head on both readout
families: a multilayer perceptron over observable features (the feature
MLP; node, goal,
context, visit counters, per-node neighbor statistics) and a hidden-state
probe on the Laya encoder (421M parameters; essentially the same
construction as the designated-position hidden-state readout of
\citet{cheng2026thisthat}). Typed decodes of the probe-family heads of
\citet{cheng2026thisthat} and \citet{ren2026jelite} run 30--65\,ms at
batch-1. Deployment is a two-tier decision stack: the readout
handles control decisions (which edge); an executor synthesizes arguments;
free-form generation stays in the LLM (as in \citealp{wu2026reflex}).

\subsection{Gated self-evolution (RSI)}
We use the term recursive self-improvement (RSI) narrowly for the
accumulate--gate--commit loop below, not for weight-space self-modification.
Decision states of executed walks are labeled with the reference next
hop---alternatively the full design's planner-oracle labels---and appended to a
replay buffer;
each round trains a candidate router on the accumulated buffer. An update is
committed only if held-out validation performance does not regress beyond
$\delta = 0.005$ (half an accuracy point); otherwise the candidate is
discarded and the incumbent kept. On the 371--380-row validation-gate fold (380 where the stream split is row-granular; App.~A), a 0.5-point bar is about 1.9 rows, comparable to the fold's own single-row resolution of 0.26--0.27 points; this $\delta$ is thus a coarse safety threshold rather than a confidence-interval-backed test. The ten-chain study's paired gate-score differences make that resolution explicit. Twelve of its 13 refusals regress the gate fold by 3 or more rows; a single 2-row refusal sits exactly at $\delta$'s own granularity (the finest distinction a 380-row gate set can support is one row); and 6 of the 27 accepted updates carry a gate-score dip worth one row that $\delta$ accepts by construction. We therefore report the 13/40 refusal count as a coarse operational safety readout, not a high-resolution statistical test of regression
(validation-gated update with rollback, cf.\ \citealp{shang2026skilaa};
held-out selection as in \citealp{nguyen2026recursiveheldout}). E4
exercises the loop under 12\% label corruption in one round: on the plotted seed-0 chain the gate score never regresses and no rollback fires there---so the guardrail is deployed and measured; across ten seeded chains its refusal path fires 13 times in 40 attempts (App.~A).

\begin{figure}[t]
\centering
\includegraphics[width=0.98\linewidth]{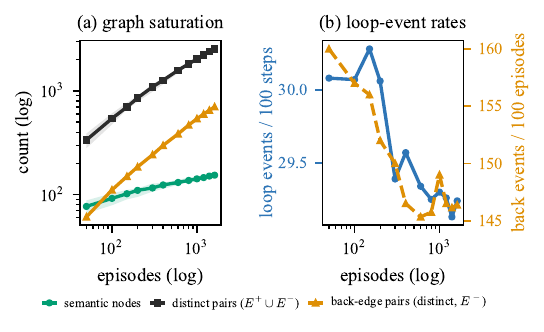}
\caption{Node-vocabulary convergence of the \graphname{} on real production traces
(subsample quantities: lines = medians of 20 without-replacement
draws at the shown size; shading = min--max over those draws, for the
node and distinct-pair series of (a) only---back-edge pairs show their
median lines). (a)~Semantic node counts saturate
quickly (77 action labels in 50-episode subsamples $\to$ 124 at 400 $\to$ 154 at the full
1{,}615), while observed transition volume (kept-frame pairs, $E^{+} \cup E^{-}$) keeps a long
tail (distinct pairs 336 at 50 episodes $\to$ 1{,}267 at 400 $\to$ 2{,}027 at 1{,}000 $\to$
2{,}536 at 1{,}615, of which 702 appear at recovery returns). (b)~Loop-event rates: fold events settle at 29.2\% of
executed steps (4{,}124 retry + 2{,}364 back) and back returns run 146 per 100 episodes---the right-axis back-event series in (b) counts these return events, 2{,}364 in full, not (a)'s 702 distinct pairs. Axes, hues and y-axis labels are figure-local
across the paper's figures (see \Cref{fig:pubsat}'s caption).}
\label{fig:convergence}
\end{figure}

\section{Experimental Setup}
\label{sec:setup}

\paragraph{Data provenance (production SOC traces).}
All experiments run on real execution logs: a Langfuse observability export
of 22{,}328 records (22{,}192 tool calls plus 136 generation records)
over 1{,}656 traces from a
production SOC agent fleet (September 2026). Keeping the trace-ID groups
with at least one tool call yields \textbf{1{,}615 episodes} over a
\textbf{154-name} tool catalog
with real per-call latencies and four domain slices (detection 1{,}174,
response 245, protection 139, recognition 54; two minor agents pool the
remaining 3 episodes). The slice names are the deployment's own agent-stage
labels and are left unmapped to external taxonomies: a slice enters the router
as part of its goal input, so we report it under its native name and do not
translate it into another vocabulary's terms.
Corpus statistics appear in \Cref{tab:corpus}. Two properties of the export
shape what can be measured and are disclosed up front: tool \emph{result}
payloads and agent chat bodies (where report text lives) are stripped, so
the router's state is reconstructed from the observable call
structure---labels and ordering; and task outcomes must be
\emph{derived}. Our derivation: an episode is \texttt{completed} when it
closes with the agent's own close-out act
\texttt{update\_goal\{status: complete\}} (present in 1{,}070 episodes,
always within the final two steps of the fold), \texttt{blocked} on
\texttt{status: blocked} (5), and \texttt{abandoned} when no close-out
occurs (540). The convention is machine-checkable (an LLM judge re-prompted
to the convention reproduces it on all 1{,}615 episodes; because the judge is handed the
convention itself, this certifies internal consistency, not independent validity) and the derived
labels are weak---convention-derived, not independently annotated---and received a manual re-reading on 30 sampled
episodes (23/3/4 across detection/response/protection), adjudicated by
an AI assistant from session logs rather than a separate human auditor
(the structural rule agrees 30/30;
an unconstrained judge prompt expecting report text inside
\texttt{update\_goal} agrees only 622/1{,}615, or 38.5\%, over the full
corpus---a shortfall that measures the export's missing
field rather than task failure: this caution is revisited in
\S\ref{sec:limitations}). Services: the re-prompted E3 opponent is
MIMO Pro V2.6 and the labeling judge is DS V4.1 Flash, both production reasoning
models behind our serving gateway (vLLM-served~\citealp{kwon2023vllm}); the typed
encoder's backbone is
Laya 0.3.21, an open non-autoregressive typed-decision engine built
on the ModernBERT-large encoder~\citep{warner2025modernbert} with a
marker-position decision head. Its
shipped per-bucket decision temperatures include an out-of-range entry, so
the zero-shot arm refits one scalar temperature on validation negative log-likelihood (henceforth NLL;
all trained arms report raw masked-renorm confidence). Service and judge latencies are
measured in vivo under production load; the zero-shot arm's 0.30\,s p50
is in-vivo CPU runtime (App.~B); typed-model timings are batch-1 GPU
wall clock (App.~B); \emph{no} latency or judge-error models are
used. The one remaining synthetic element is deliberate and scoped: E4's
12\% label corruption is a controlled stress injection (a stand-in for
policy regressions), not a claim about the deployment's noise rate.

\begin{table}[t]
\centering
\caption{Real SOC corpus after loop folding (\S\ref{sec:method}): 22{,}192
tool calls grouped into 1{,}615 episodes over 154 actions. ``decision
steps'' count advance (hop) steps; the router's decision states add the
kept recovery returns (train/val/test $+$1{,}684/$+$186/$+$494 $=$ 2{,}364),
giving the 18{,}068 rows ($12{,}613{+}1{,}580{+}3{,}875$) all SOC
experiments draw on. ``loop events'' count executed retry and back fold events
(4{,}124 retries, which leave the decision stream, plus 2{,}364 returns,
which stay).
``completed'' is the derived
close-out outcome (\S\ref{sec:setup}). The self-evolution stream and its
rollback gate for E4 are carved from the validation fold.}
\label{tab:corpus}
\vspace{1mm}
{\footnotesize
\begin{tabular}{lrrrr}\toprule
split & episodes & completed (\%) & decision steps & loop events \\\midrule
train & 1,125 & 67.3 & 10,929 & 4,615 \\
val & 145 & 60.0 & 1,394 & 519 \\
test & 345 & 65.5 & 3,381 & 1,354 \\
all & 1,615 & 66.3 & 15,704 & 6,488 \\
\bottomrule\end{tabular}
}
\end{table}

\paragraph{Replay protocol (held-out logs as the environment).}
The production logs \emph{are} the environment: held-out episodes are
replayed decision-by-decision and every policy faces the same reference
behavior---the deployed fleet's own choices at each realized state. Per
decision the policy sees the declared option set (the current
label's outgoing edges in the train-induced graph, plus
\texttt{stop}) and a prefix rendering that differs by arm. The feature
arm summarizes the folded kept prefix, while the probe and zero-shot
state texts render an eight-entry window of the raw label list, cut at
the kept frame's fold ordinal (its rank among kept frames), which serves as the raw-list index. Each
entry is rendered by its name tail (the substring after the last \texttt{\_\_}, or the
whole entry when it contains none), so distinct catalog actions can
collide (two \texttt{search} actions surface identically). The
zero-shot arm's option criteria render the same tails with underscores
as spaces, barring the \texttt{<stop>} criterion whose text reads
\texttt{conclude}; colliding options then present identical choice texts. The
window runs behind the
current label by exactly the retries preceding the frame (the retries
stay in the list; on such rows the typed arms never see the current
label), while the LLM prompt renders the same window with full labels
(not name tails) and appends the current label explicitly. Scoring is teacher-forced against the logged next hop; episodes are
scored two ways---strict survival (an episode terminates at its first
divergence) and the wrong-hop gap accumulated over the whole episode
(cost conventions 1 and 3 per wrong hop are both reported throughout). Because the reference is a realized
trajectory rather than an oracle, task success inherits the derived outcome
labels above, and the \emph{declared option space is itself scored}: 7.9\%
of test targets lie outside the train-declared options
(declared-space coverage 92.1\%, or 91.3\% when recovery edges are restricted to
states the running prefix has already visited---29 rows, every one of them a
return to an unvisited state, with the mean space falling from 40.2 to 39.2
options and the verified arms' masked accuracy moving by at most 0.1 points; of those 307 out-of-space targets, 274 are
SOC-service names and 33 are tool names outside the catalog), an arithmetic upper bound no readout inside the
declared space can exceed---we report accuracy both overall and within the
declared space. The carve is deterministic and trace-disjoint: every trace
hashes to train/validation/test at 70/10/20 (first 8 hex digits of its
\texttt{sha256}, mod 100), so every train-versus-held-out claim below
except the E4 gate comparisons refers to episodes whose traces no fit
saw (the self-evolution stream's finer split
of the validation fold is decision-row-granular: rows, not traces, are
split, a leakage risk disclosed here and now measured---re-cutting the
stream and the gate at whole-trace granularity moves held-out accuracy by under
0.1 points and leaves every E4 conclusion in place).

\paragraph{Baselines and arms.}
E2 compares a per-node majority arm (raw per-node argmax) against
multinomial logistic regression and typed readouts trained on one
observable-state feature set (node, goal, prefix statistics, visit
counters, per-node neighbor statistics); the typed readouts are trained with cross-entropy versus the
CE-plus-Brier variant (5 training seeds each); in the
edge-ID variant the readout space is the induced graph's
edge identifiers rather than target-action classes. Two class-space readouts use
Laya: zero-shot typed choice over each state's declared
options, and a frozen-encoder hidden-state probe---a linear head on the
encoder's designated readout position---trained with the same CE
and CE-plus-Brier objectives. E3 compares: a re-prompted MIMO Pro V2.6 stand-in over the reconstructed
state (MIMO Pro V2.6, a production reasoning model at our serving gateway---one
call per decision, without the fleet's per-stage orchestration and
without tool results---not the deployment's status quo) deciding every
hop from the same reconstructed state and options (under the richer
full-label prompt rendering of the replay protocol); typed readouts acting
greedily within the masked option set; and typed with accept/escalate at
$\tau \in \{0.55, 0.75, 0.9\}$, which routes low-confidence states to
MIMO Pro V2.6. Readouts are named after their table rows: \emph{feature MLP} scores
engine-derived stream statistics, \emph{hidden-state probe} is a probe fit on Laya's
frozen hidden states, and \emph{typed encoder} is the Laya backbone itself scoring the
declared options.

\paragraph{Metrics and protocol.}
Per decision: top-1 agreement with the logged hop, masked-renormalized over
the declared options for the trained arms, reported overall and within
the declared space (for the baseline rows the displayed cells are raw
top-1; their option-masked variant merely gates the raw argmax on
landing in the option set---no renormalization);
calibration (ECE-15, per the \S\ref{sec:method} convention; it is computed on the 3{,}568 in-space rows for the
zero-shot arm---a typed-choice head gives no score to an out-of-space
target---and on all 3{,}875 rows with out-of-space targets counted wrong for the
other arms, whose full-vocabulary heads still define a masked-renorm
confidence there). Risk--coverage curves~\citep{geifman2017selective} and the area under the
risk--coverage curve (AURC) for the accept/escalate
gate complete the per-decision set. Per episode: strict survival (no divergence before the log ends),
LLM calls per decision, and the wrong-hop gap accumulated over the whole
teacher-forced episode under both per-hop cost conventions (1 and 3). Difficulty profile: domain (agent-stage) slices
and local branch-factor buckets ($\le$3, 4--10, 11+: the train-graph
out-degree at the current node, stop options excluded).
Uncertainty: 5 training seeds for the readout arms (mean $\pm$ std); the E3 replay and the E4 refit chain are single-seed (seed 0) sequences (E4's ten-chain stress study is
aggregated at the unit of one refit chain: 30.3\% and 31.0\% are across-chain
means of start and end held-out accuracy, $\pm 0.5$ points---one standard
deviation across the ten chains' end accuracies);
service and judge latencies are in-vivo wall-clock (p50/p90), the
zero-shot arm's runtime in-vivo CPU (App.~B), typed-model timings
batch-1 GPU wall clock (App.~B). Each of E1--E4 is annotated
with the claim it tests.

\section{Results}
\label{sec:results}

\paragraph{E1: the induced action graph's node vocabulary converges (its edge set runs a long tail; claim \pnum{1}).}
\Cref{fig:convergence} provides quantified saturation curves on real
production traces, adding the transition-pair long tail and the loop-event
mass to the usual state-count view. Across 20 without-replacement subsamples per size (medians quoted
here and in \mbox{\Cref{fig:convergence}}), the
semantic node vocabulary saturates fast: 77 action labels appear in
50-episode subsamples, 124 at 400, 142 at 1{,}000, against 154 in the full
corpus. Transition discovery is the honest counterpoint---observed
transition pairs keep a long tail (336 at 50 episodes $\to$ 1{,}267 at 400
$\to$ 2{,}027 at 1{,}000 $\to$ 2{,}536 at 1{,}615), and the recovery side
never closes either: 702 pairs appear at recovery returns, 481 of them
doubling as forward transfers in other contexts
(\S\ref{sec:problem}). Loop mass is heavy and stable: 29.2\% of executed steps fold
into loop events (4{,}124 retry---63.6\% of loop events, in 2{,}164 bursts---and
2{,}364 back returns---36.4\%, 146 per 100 episodes; revisit distances
concentrate at 1--3 steps behind the current frame, with 681 taking the
immediate-repeat distance of one). Two consequences: the router's \emph{label space} closes
early---the stability pattern of \citet{cho2026automata} recurs in the
node curve (qualitatively, a small saturating node set of 77 to 154
labels rather than their 7--43 states; \mbox{\Cref{fig:convergence}a}),
now alongside the transition-pair and loop-mass axes---while per-node
option sets keep thickening. The routing task on
real traces is therefore bounded-vocabulary but long-tailed option
selection: exactly the regime where a masked, calibrated typed readout
constrains the damage of open-vocabulary guessing.

\paragraph{E2: the router reaches top-1 agreement with the fleet (32.8\%) and decides in milliseconds (claim \pnum{2}).}
\begin{table}[t]
\centering
\caption{Next-hop agreement on the held-out test decisions: top-1
accuracy over 3{,}875
decision states; for the trained readout arms, masked top-1
renormalized over each state's declared options; raw top-1 for the
majority and logistic rows (option-masked accuracy is computed for both baseline rows on the
public streams and for the majority arm on all nine streams, App.~B;
it equals raw for the majority arm everywhere, while the logistic pair
diverges on the two ToolBench streams~\citealp{qin2023toolllm}: 35.7\% masked versus 36.4\% raw
at 7{,}318 names and 34.8\% masked versus 36.0\% raw on the 351-name
evaluation subset, \S\ref{sec:results}). Uncertainty: 5 training seeds for the readout
arms, $\pm$ std---accuracy fractions, e.g.\ $0.004$---in the note column. ``in-space'' restricts to the 92.1\% of decisions whose logged target lies
inside the declared option space. ECE-15 and AURC are probabilities printed to
three decimals. ``--'' = undefined (the majority and logistic rows
carry no calibration cells, no in-space cells, and no AURC entry: the majority arm has no
confidence ranking and no calibrated distribution is fitted for the logistic row; AURC is likewise not computed for the
probe and zero-shot rows---their risk--coverage curves come only from coarse
7-point grids); note-column ``--'' = no seed spread and no config
entry (the zero-shot row reports its fitted decoding temperature $T$
instead). Probe and zero-shot fitting details are in
App.~B. $^{\dagger}$The paired interval against the indicated reference
includes zero (\Cref{tab:paired}): the ordering among the marked rows is a point
estimate, not a separated difference.}
\label{tab:router}
\vspace{1mm}
{\footnotesize
\resizebox{\linewidth}{!}{{\setlength{\tabcolsep}{4.5pt}\begin{tabular}{lrrrrl}\toprule
readout arm & acc.\ (\%) & in-space (\%) & ECE-15 & AURC & $\pm$ std / config \\\midrule
majority arm & 26.8 & -- & -- & -- & -- \\
logistic regression$^{\dagger}$ & 32.3 & -- & -- & -- & -- \\
feature MLP, CE$^{\dagger}$ & 32.8 & 35.6 & 0.045 & 0.409 & $\pm 0.001$ \\
feature MLP, CE$+$Brier$^{\dagger}$ & 32.7 & 35.5 & 0.048 & 0.409 & $\pm 0.003$ \\
edge-ID readout, CE$^{\dagger}$ & 31.4 & 34.1 & 0.082 & 0.432 & $\pm 0.007$ \\
edge-ID readout, CE$+$Brier$^{\dagger}$ & 31.6 & 34.3 & 0.080 & 0.431 & $\pm 0.004$ \\
hidden-state probe, CE & 30.3 & 32.9 & 0.123 & -- & $\pm 0.007$ \\
hidden-state probe, CE$+$Brier & 30.7 & 33.3 & 0.122 & -- & $\pm 0.004$ \\
hidden-state probe (edge-ID), CE$+$Brier & 26.6 & 28.9 & 0.458 & -- & $\pm 0.004$ \\
Laya zero-shot typed choice & 8.6 & 9.3 & 0.031 & -- & $T=3.0$ \\
\bottomrule\end{tabular}}
}}
\end{table}
\Cref{tab:paired} replaces the ordering reading with paired tests on the same
held-out states and 95\% intervals from 2{,}000 trace-clustered bootstrap
resamples over the 345 test episodes (resampling decisions independently would
understate the interval: dependence inside an episode is strong). Two of the six
comparisons separate---the readout against the per-node majority arm
($+5.7$ points, CI $[+4.6,+6.9]$) and the edge-ID collapse on the hidden-state
probe ($-3.8$ points, CI $[-5.0,-2.6]$)---while the CE/CE$+$Brier ordering, the
feature MLP against logistic regression, and the edge-ID penalty \emph{on the
feature arms} all have intervals that include zero and are therefore reported as
point estimates. \Cref{tab:buckets} gives the branch-factor view: the headline
accuracy is a weighted average dominated by the $11+$ bucket (89.6\% of rows),
the $\le 3$ bucket is effectively solved ($92\%$ majority, $99\%$ in-space for
the best arm), and the small middle bucket is the hardest in-space case.
Finally, the information budget behind these numbers is small: once node identity
and the eight state statistics are given, adding seven prefix-latency features
buys $+0.4$ points and adding argument-shape counts on top of them $+1.3$ points
(five seeds, \texttt{feat\_lat\_results.json}, Reproducibility), so the ordering
among feature arms should not be read as evidence that richer signals were the
missing ingredient.

\begin{table}[t]
\centering
\caption{Paired tests on the same held-out states (seed-0 fits;
\texttt{e10\_paired\_ci.json}, \texttt{e11\_probe\_edge\_ci.json}). $\Delta$ is in
accuracy points with a 95\% CI from 2{,}000 trace-clustered bootstrap resamples;
``discordant'' gives the counts (first arm right / second arm right). Rows read
\emph{point estimate only} have a CI that includes zero: their ordering in
\Cref{tab:router} is a point estimate, not a separated difference.}
\label{tab:paired}
\vspace{1mm}
{\footnotesize
\resizebox{\linewidth}{!}{{\setlength{\tabcolsep}{4.5pt}\begin{tabular}{lrrrl}\toprule
paired comparison (same states) & $\Delta$ (pts) & 95\% clustered CI & discordant & reading \\\midrule
feature MLP, CE$+$Brier $-$ CE & $-0.13$ & $[-0.53,\ +0.26]$ & 35 / 30 & point estimate only \\
feature MLP, CE$+$Brier $-$ majority arm & $+5.70$ & $[+4.60,\ +6.90]$ & 105 / 326 & separates \\
feature MLP, CE$+$Brier $-$ logistic regression & $+0.23$ & $[-0.48,\ +0.93]$ & 101 / 110 & point estimate only \\
edge-ID $-$ class space (features, CE$+$Brier) & $-0.83$ & $[-1.67,\ +0.03]$ & 176 / 144 & point estimate only \\
edge-ID $-$ class space (probe, CE$+$Brier) & $-3.79$ & $[-5.01,\ -2.62]$ & 212 / 359 & separates \\
edge-ID CE$+$Brier $-$ edge-ID logistic & $-0.15$ & $[-1.05,\ +0.76]$ & 176 / 170 & point estimate only \\
\bottomrule\end{tabular}}
}}
\end{table}

\begin{table}[t]
\centering
\caption{The same decisions split by the branch factor $|A(v)|$ at the state
(3{,}875 test decisions). Only 3.6\% of states fall in the middle bucket, so the
headline is dominated by the $11+$ bucket (89.6\% of rows). ``majority'' is the
per-node majority-vote arm; MLP and edge-ID are the feature arms, masked accuracy
with in-space accuracy in parentheses; mean $|A(v)|$ is over declared options
(\texttt{stop} included). \texttt{e8\_slices.json} additionally carries a
pooled train$+$val$+$test \texttt{by\_outdeg} table with the same accuracy
magnitudes over 19{,}375 rows; the counts and accuracies quoted here are
test-split, and our seed-0 probe refit reproduces them (92.0/22.3/26.3 against
the five-seed 92.2/21.2/25.9 of \S\ref{sec:results}).}
\label{tab:buckets}
\vspace{1mm}
{\footnotesize
\resizebox{\linewidth}{!}{{\setlength{\tabcolsep}{4.5pt}\begin{tabular}{lrrrrrr}\toprule
bucket ($|A(v)|$ of the state) & $n$ & share & mean $|A(v)|$ & majority & MLP CE$+$Brier & edge-ID \\\midrule
$\leq 3$ & 263 & 6.8\% & 1.23 & 92.0 & 93.9 (99.2) & 93.2 \\
$4$--$10$ & 139 & 3.6\% & 7.90 & 34.5 & 38.9 (54.0) & 36.7 \\
$11+$ & 3{,}473 & 89.6\% & 44.46 & 21.5 & 27.6 (29.8) & 26.8 \\
\bottomrule\end{tabular}}
}}
\end{table}
\begin{figure}[t]
\centering
\includegraphics[width=0.98\linewidth]{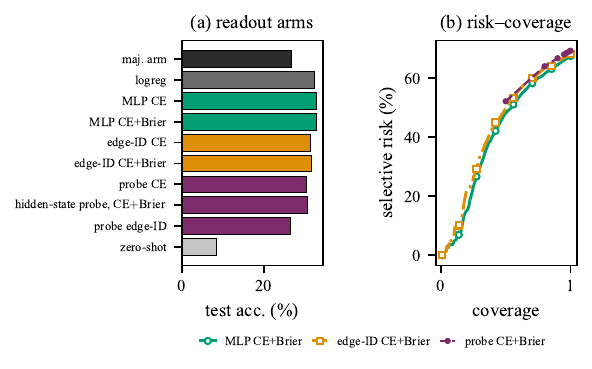}
\caption{(a)~Readout arms on the held-out test decisions (masked top-1;
raw top-1 for the majority/logistic rows; the zero-shot row counts
out-of-space targets as errors, as the masked rows and
the majority arm
do). (b)~Risk--coverage of the CE-plus-Brier arms (masked-renormalized
distributions; the edge-ID probe arm is not drawn): confidence
separates risk sharply at the top of the ranking in the feature arms and
thinly across the bulk in both families (the probe grid starts at 50\%
coverage). Dense curves live in
\texttt{real/e2\_results.json} (\texttt{curve\_mean}; seed 0 for the
feature arms); the probe arm is
drawn from its seven-point, five-seed-mean risk@coverage grid in
\texttt{real/e2\_layaprobe\_results.json}.}
\label{fig:router}
\end{figure}
\Cref{tab:router} states the difficulty plainly: next-hop prediction
on production traces tops out at 32.8\% agreement on the state features (feature MLP, CE;
35.6\% within the declared space) with its CE-plus-Brier twin at 32.7\%,
the logistic-regression arm at 32.3\% (raw top-1) and the majority arm at 26.8\%. These absolute levels measure the task as much
as the models: the reference behavior is a realized fleet policy rather
than a planner oracle, states offer 40.2 declared options on average
(median 46,
\texttt{stop} included), and 7.9\% of targets fall outside the
train-declared option space (a 92.1\% declared-space arithmetic upper bound, 91.3\% under prefix-marked reachability, for masked
readouts; that restriction shifts the feature-MLP and edge-ID arms by at most $0.0005$ accuracy and
leaves 29 of 3{,}875 rows without their target, so no reported level depends on it). Three structural findings hold on production traces. At this catalog size semantic class sharing
beats structural book-keeping: the class-space readout beats the
edge-ID readout on shallow features (32.7\% versus 31.6\% masked; both CE$+$Brier, 5 seeds)
and by far on the hidden-state probe (30.7\% versus 26.6\%; the edge-ID
probe's ECE explodes to 0.458)---the reverse of our design bet, discussed
in \S\ref{sec:limitations}. Also, the CE-plus-Brier
objective pays where the readout trails and is flat where it
already leads (the probe trails here and gains; the feature leads and is flat): on the feature MLP it gives up 0.05 accuracy
points and worsens ECE by 0.002 versus plain CE (32.7\% versus 32.8\%, ECE 0.048 versus
0.045, deltas on unrounded 5-seed means), while
on the hidden-state probe it adds 0.4 points (30.3\% $\to$ 30.7\%) and
cuts the seed spread ($\pm 0.007$ $\to$ $\pm 0.004$). And
the designated-position hidden-state readout of \citet{cheng2026thisthat}
is its system's answer path and wins system-level comparisons; here, encoder
held fixed, the structured feature arm beats the designated
hidden-position probe (32.7\% versus
30.7\%; in-space 35.5\% versus 33.3\%). For a proprietary 154-name catalog
described by name only, the structured feature arm wins this
comparison (although representation, readout family and training
budget are confounded here: the features summarize the folded kept
prefix while the probe's state text renders raw logged labels through
the name-tail surface of \S\ref{sec:setup} and trails the current label
wherever retries precede it, by up to dozens of raw steps; the probe
reads through a linear head on encoder state where the feature arm is
an MLP over the observable features; and the probe's reported fit runs
90 epochs against the feature arm's 60). Zero-shot typed choice (8.6\% overall, 9.3\% within
the declared space; fitted temperature $T{=}3.0$, the ceiling of an
eight-point refit grid $T\in\{0.5,0.7,1.0,1.3,1.6,2.0,2.5,3.0\}$ and still the
ceiling when the grid is extended to $T{=}1000$: the raw head is over-confident
and the fit is unbounded (validation NLL falls monotonically to the
uniform-prediction limit $3.40$; the argmax is invariant, the risk--coverage
curve moves by at most $0.0006$, and ECE-15 is grid-dependent---$0.031$ at
$T{=}3.0$ against $0.010$ at $T{=}1000$---so this arm's calibration is not to be
read as calibrated confidence) shows what transfer alone
buys. Calibration is another honest limit: the probe's ECE sits at
0.122 against 0.045 for the best feature arm, and confidence on the feature arms
stratifies sharply only at the top of its ranking---selective risk on
features falls to 3\% at 5\% coverage and 23\% at the top quartile (the seed-0 dense curve; full risk--coverage curves in \Cref{fig:router}),
then rises more slowly (48\% at 50\% coverage to 67\% at full
coverage on the feature MLP with CE$+$Brier, five-seed means; 52\% $\to$ 69\% at the same
two anchors on the hidden-state probe with CE$+$Brier). The accept/escalate gate therefore
lives or dies with the hybrid sweep in E3. Where the data are easy the readout is easy: the hidden-state
probe's branch-factor and domain-slice refits (60-epoch fits over its frozen encoder features,
App.~B; \texttt{real/e8\_slices.json}) answer the 263 test
states in the $\le$3 local-branch-factor bucket at 92.2\%
(aggregated over 5 seeds $\times$ the test split), 21.2\% on the 139
states in the 4--10 bucket, and 25.9\% on the 3{,}473 states in the 11+
bucket---the middle bucket dips below the largest. By
domain slice the same probe refits average 28.9\%
(detection), 41.9\% (response), 29.3\% (protection) and 42.2\%
(recognition)---an ordering \S\ref{sec:limitations} cautions against reading
directly, since stage accuracy conflates routing with pipeline role
(closure rates in App.~A).

\paragraph{E3: end-to-end replay against the re-prompted MIMO Pro V2.6 stand-in (claim \pnum{3}).}
\begin{table}[t]
\centering
\caption{Held-out replay on the 345 test episodes (3{,}875 decisions,
\S\ref{sec:setup}). All policies share the declared option sets and the
fleet reference behavior. Match = decision-level top-1 agreement with the
logged hop (teacher-forced); survival = episodes that replay with no
divergence before the log ends (the episode dies at its first
divergence); gap = wrong-hop cost per episode---every diverging hop of
the whole teacher-forced episode is charged---under conventions
$c_1$ (1/wrong hop) and $c_3$ (3/wrong hop); LLM calls/dec.\ = escalation
rate; every typed and hybrid row replays one probe fit (CE$+$Brier,
seed 0). Gap columns come from one wrong-hop count each: $c_3 = 3c_1$
exactly, and displayed entries round independently.}
\label{tab:endtoend}
\vspace{1mm}
{\footnotesize
\begin{tabular}{lrrrrr}\toprule
policy & match (\%) & survival (\%) & gap $c_1$ & gap $c_3$ & LLM calls/dec.\ (\%) \\\midrule
MIMO Pro V2.6 every hop & 9.6 & 0.0 & 10.157 & 30.470 & 100.0 \\
hidden-state probe, CE+Brier & 30.7 & 0.6 & 7.788 & 23.365 & 0.0 \\
typed+esc @ $\tau{=}0.55$ & 20.9 & 0.0 & 8.890 & 26.670 & 74.1 \\
typed+esc @ $\tau{=}0.75$ & 19.1 & 0.0 & 9.084 & 27.252 & 80.2 \\
typed+esc @ $\tau{=}0.9$ & 16.5 & 0.0 & 9.377 & 28.130 & 85.0 \\
\bottomrule\end{tabular}
}
\end{table}
\begin{figure}[t]
\centering
\includegraphics[width=0.98\linewidth]{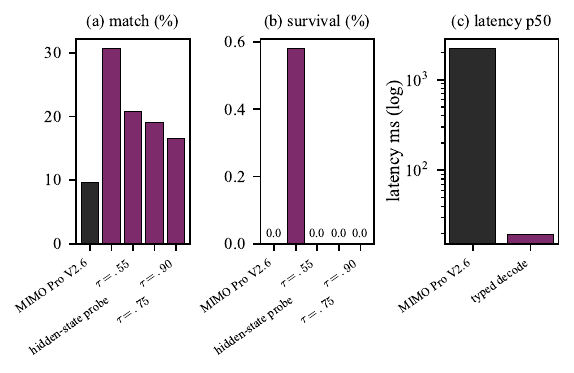}
\caption{End-to-end replay (\Cref{tab:endtoend}). (a)~Decision match;
(b)~Episode survival under strict divergence; (c)~Decision
latency: MIMO Pro V2.6 in-vivo p50 2.2\,s versus 19.5\,ms p50 for the typed
decode's GPU typed-choice head (full typed-choice predict, encoder pass
included---the typed path's upper price; the E3 replay's probe rows
decode at the 10\,ms marginal encoder pass: App.~B). The accept/escalate rows spend LLM calls only where the
readout is least confident---and score strictly below the pure hidden-state-probe row; on this
stream the escalated hops carry the opponent's low agreement into the route,
which is consistent with the loss but is not isolated by our design
(\S\ref{sec:results}). The typed-family bars in (a) and (b) come from one
hidden-state probe fit (CE$+$Brier, seed 0; probe family drawn purple);
the purple bar in (c) is the typed path's own decode price.}
\label{fig:endtoend}
\end{figure}
\Cref{tab:endtoend} delivers the end-to-end outcome
(\Cref{fig:endtoend}). With all 3{,}875 decisions of the 345 held-out
episodes scored against the fleet's own behavior, the re-prompted MIMO Pro V2.6 stand-in (one
reasoning call per hop; its free-text replies are matched to option names
post hoc, 66 replies---1.7\%---unparsed and scored as divergences)
agrees with the reference at 9.6\%; the hidden-state probe holds 30.7\% at
zero LLM calls; and the accept/escalate hybrids land \emph{below} the
typed layer---20.9\% / 19.1\% / 16.5\% at $\tau \in \{0.55, 0.75, 0.9\}$
under escalation rates of 74.1\% / 80.2\% / 85.0\% (MIMO Pro V2.6 consulted on
the readout's least-confident states; the $\tau$ sweep traces the
agreement/cost curve that slopes down). The episode-level
measures agree with the per-decision read: strict survival keeps 2 of 345
episodes for the hidden-state probe (0.6\%) and none for the opponent or the
hybrids---any-divergence death on ten-plus-hop episodes asks for
perfection in every hop---while the wrong-hop gap charges 10.2 wrong
hops per episode against the opponent (30.5 under $c_3$) and 7.8 (23.4)
for the hidden-state probe. The hybrids land at 8.9/9.1/9.4 (26.7/27.3/28.1 under
$c_3$). The cost asymmetry holds regardless of who
wins the agreement column: every typed decision is one bounded readout
pass---a typed-choice predict at 19.5\,ms p50 with the encoder pass
included (17 declared options measured, below the 40.2-option held-out mean), or, on the replay's probe
rows, that pass alone (10\,ms) plus
a linear head (App.~B)---while every MIMO Pro V2.6 decision is a reasoning generation
at 2.2\,s p50 (7.4\,s p90) under the serving configuration's 2{,}048-token completion budget.
We account for cost on the three axes the export actually carries---decision
latency, LLM calls per decision, and the wrong-hop cost of a divergence---and not
in tokens or currency: the stripped stream we work from records no per-call
usage, so a token-level comparison here would be modelled, not measured.
Two readings follow. Matching this fleet is hard for everyone---the
reference is a realized multi-agent policy whose tool \emph{results} our
export strips (\S\ref{sec:limitations})---and within that hard task the
cheap readout beats the zero-shot opponent outright, $3.2\times$
its agreement at two orders of magnitude
less decision time. This comparison is asymmetric in construction: the readout is fit on 12{,}613 logged reference decisions while the stand-in is only ever re-prompted zero-shot with tool results stripped (66 of its replies, 1.7\%, fail to parse into a valid tool choice and score 0 against the fleet's logged next hop), so 9.6\% is a transfer-floor number for a stripped stand-in, not the deployed component's own capability ceiling. Tuning is the difference: jpt-4b, LoRA-tuned for one epoch on the training fold, reaches 31.9\% against the typed encoder's 32.2\% on the same held-out decisions, a gap an exact paired test does not resolve ($p=0.68$, 229 decisions the tuned model gets and the typed encoder misses, 239 the other way), at 244\,ms per decision against 10\,ms (\Cref{tab:e23scale}). The hybrid
result is a negative one for confidence-gated cascade calibration on
this stream. The gate
routes roughly four of every five hops to the LLM (3.7--4.3 of every five hops at
74.1/80.2/85.0\% for the three $\tau$ thresholds) and inherits the opponent's
low agreement on those hops, so rather than beating both parents on
their complementary states it
sits strictly between its parents---an escalation rule needs per-state
competence, not just confidence, before it can pay for itself here.

\paragraph{Threshold scale, and where routes first diverge.} The raw thresholds
above are not scale-free: $p_{\max}$ lives on a per-node simplex whose dimension
is the option count, and the threshold that matches a given escalation rate
correlates with $1/|A(v)|$ at Pearson $r=0.52$. Re-running the sweep on an
excess-over-uniform scale, an entropy-normalized scale, the top-1/top-2 margin,
and a threshold-free rule that escalates the least-confident fraction, at
\emph{matched} escalation rates, leaves the reading unchanged: at 74.1\%
escalation the hybrid reaches 0.2085 (raw) / 0.2263 (excess over uniform) /
0.2098 (entropy-normalized) / 0.2101 (margin) / 0.2085 (rank only) against
0.3061 for the pure hidden-state-probe row, so the cascade's loss is not an artifact of an
uncalibrated threshold and no tested scale closes more than 1.8 of the 9.8-point
gap. Where a route dies is a quantity the per-decision mismatches do not carry:
of the 343 episodes that diverge at all under the hidden-state probe, 60 diverge at
the first hop and 185 at the second (median 2)---routes are lost on the way in;
331 of those first divergences sit at states with $11+$ declared options, 12 at
out-of-space targets, and 258 fall on hops the hybrid had escalated.

\paragraph{E4: the gated loop absorbs the corruption stress on average, and the rollback gate is exercised---13 of 40 candidate updates refused over ten seeded chains (claim \pnum{3}).}
\begin{figure}[t]
\centering
\includegraphics[width=0.55\linewidth]{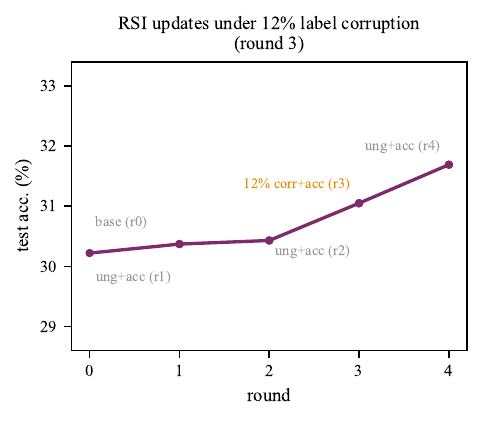}
\caption{Held-out \emph{test} decision accuracy across the self-evolution stream (a
seeded random 1{,}200-row subset of the validation fold, permutation
seed 7, played as 300-decision update batches; the 380 validation rows left
out of the 1{,}200-row draw, from the 1{,}580-row validation fold, are the gate set). The round-3 batch carries 12\% label
corruption (App.~A); on the seed-0 row-level chain plotted here held-out
accuracy never falls across refits, rises 30.2\% $\to$ 31.7\%, and holds
through the corrupted one (the flips are
about 0.3\% of the 12{,}613-decision base). The gate set governs
commits: on this chain the gate score never regresses beyond the gate tolerance
$\delta{=}0.005$. Across ten seeded chains (row- and trace-level stream splits;
five chains per mode over shared seed offsets, App.~A) held-out accuracy
averages 30.3\% $\to$ 31.0\% ($\pm 0.5$ points) and the gate refuses 13 of the 40
candidate updates (6 of the 10 chains refuse at least one), so the single
chain under-reports the guardrail's workload (point
tags concatenate the round's events with \texttt{+} and carry the
round number: \texttt{ung+acc (r1)} names the raw refit \emph{ungated}
update and the gated commit \emph{accept} at round 1,
\texttt{12\% corr+acc (r3)} the injected round, \texttt{base (r0)} the
base fit). The test
split is never trained on.}
\label{fig:rsi}
\end{figure}
Across four rounds (\Cref{fig:rsi}) the gated loop accumulates the
self-evolution stream in 300-decision batches---each candidate is a refit of
the hidden-state probe on accumulated data (train split plus accepted
batches). The stress arrives at round
3 (12\% of the batch's labels flipped to uniformly drawn non-target
options, App.~A): against a 12{,}613-decision base the injected flips do
not dent held-out accuracy---the five fits run 30.2\% $\to$ 30.4\%
$\to$ 30.4\% $\to$ 31.0\% $\to$ 31.7\% on the untouched test split with
the gate score rising alongside (25.0\% $\to$ 27.1\%)---and every refit,
corrupted one included, is committed (5 fits on the chain, base plus
four refits; 4 of 4 candidate updates accepted; 0 reverts at
$\delta{=}0.005$). What the run demonstrates is monotone accumulation
and noise absorption at this injection rate, not a gate rescue: a stress
sized to bite was run as that honest next test, and the answer is partly
negative. At 30\% and 50\% (four chains each, same recipe, injection still in
round 3) the gate refuses 7 and 8 of 16 candidate updates against 8 at 12\%
while the corrupted update's gate score falls further
($0.0013$ and $0.0059$ mean drop against $0.0007$ at 12\%), and final held-out
accuracy stays within $0.002$ of the base ($31.2\%$ and $30.9\%$ against
$30.8\%$)---so the refusal path is exercised at every rate but neither the
refusal count nor the held-out metric tracks the injection size, because the
$0.5$-point threshold already separates the same updates. The other half of
that test---corruption against a smaller base---was run too: at a quarter of
the base ($3{,}153$ decisions) the fit starts $2.6$ points lower ($27.6\%$
against $30.3\%$), the gated loop recovers most of the gap over the four
updates ($29.5\%$ and $29.6\%$ against $30.8\%$ and $31.3\%$), and the rollback
path fires on every chain ($3/2/1$ refusals at 12\% and $3/0/1$ at 30\%),
including the chain that never fires at the full base---so base size, not
injection size, is what makes the gate bite. Direction check
at the same 12\% rate: a second batch replaces the \emph{observation}
instead---the feature row shown to the model is taken from another stream row
with a different logged target, the label left intact---and is absorbed the same
way (6 rollbacks across the same five runs against 7; final held-out accuracy
31.08\% against 31.05\%), so the loop's behaviour here does not depend on which
side of the imitation pair is corrupted, and the gate is not a mislabel-specific
detector. Richer
world- and harness-evolution loops are
complementary~\citep{xue2026sciencebuddy,zheng2026dreamrsi,neo2026neohorse}.

\paragraph{Ablations (A1--A3).}
\begin{table}[t]
\centering
\caption{Ablation summary (A2 cells: masked acc.\ / ECE-15; 5 seeds
where trained; A1 and A3 rows are counts and rates). A1: what loop folding removes from the decision stream. A2:
loss ablation (arrows run CE $\to$ CE$+$Brier) on the readout arms. A3:
accept/escalate threshold sweep---summary row here, full hybrid rows in
\Cref{tab:endtoend}, risk--coverage curves in \Cref{fig:router}.}
\label{tab:ablation}
\vspace{1mm}
{\footnotesize\raggedright
\begin{tabular}{ll>{\raggedright\arraybackslash}p{0.55\linewidth}}\toprule
 & item & measured effect (acc.\ /\ ECE-15 where two-sided) \\\midrule
A1 & loop folding & 22,192 positions $\to$ 18,068 decision states (15,704 advances + 2,364 kept returns); 4,124 retries dropped, 6,488 loop events tagged (29.2\% of steps) \\
A2 & feature MLP & 32.8\,/0.045 $\to$ 32.7\,/0.048 \\
 & edge-ID readout & 31.4\,/0.082 $\to$ 31.6\,/0.080 \\
 & hidden-state probe & 30.3\,/0.123 $\to$ 30.7\,/0.122 \\
A3 & escalation sweep & $\tau{=}0.55/0.75/0.9$ $\to$ 74.1/80.2/85.0\% LLM calls, match 20.9/19.1/16.5\% \\
\bottomrule\end{tabular}
}
\end{table}
(A1)~Without loop folding the readout would face the raw stream of
22{,}192 positions of which 29.2\% are loop bookkeeping: folding removes
only the 4{,}124 retries (18{,}068 decision states remain: 15{,}704
forward plus 2{,}364 recovery returns---the same accounting E5's
237-kept-frame example spells out) and compresses the state-ID space to
154 semantic labels
(\mbox{\Cref{fig:convergence}a}). (A2)~CE$+$Brier training
is flat against plain CE on the class-space readout (paired $\Delta=-0.13$
points, 95\% trace-clustered CI $[-0.53,+0.26]$: an unresolved ordering, not a
win; \Cref{tab:paired}) and nudges
the edge-ID readout up a notch (31.4\% $\to$ 31.6\%, ECE 0.082 $\to$
0.080); it buys its margin on the hidden-state probe (the E2 deltas,
tabulated in \Cref{tab:ablation}): a hedge for whichever readout
trails rather than free accuracy. (A3)~The accept/escalate sweep lives in the hybrid rows of
\Cref{tab:endtoend} (escalation rate is monotone in $\tau$ by
construction) and in the risk--coverage curves of \Cref{fig:router};
the deployed gate thresholds the masked-renormalized distribution, and
out-of-space emissions are masked back into range (\S\ref{sec:setup}).

\paragraph{E5: external replication on public corpora (generalization probe).}
\begin{figure}[t]
\centering
\includegraphics[width=\linewidth]{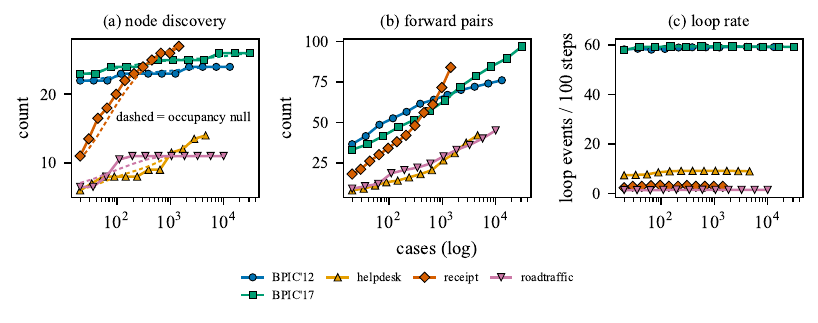}
\caption{Saturation on the five process logs that carry $\geq$20 cases
(statechart~\citealp{leemans2018statechart} is a single 17{,}912-event session and cannot support a
subsampling curve; its census sits in \Cref{tab:pub}). Twelve
without-replacement subsample sizes $\times$ 20 draws each; p50 curves.
(a)~Distinct-node discovery climbs toward each catalog ceiling;
receipt's 27th and helpdesk's last label appear only at their case
boundaries~\citep{buijs2014receipt,polato2017helpdesk}. (b)~First-visit forward pairs ($E^{+}$ only; the SOC volume in
\mbox{\Cref{fig:convergence}} counts kept-frame unions $E^{+} \cup
E^{-}$) are still climbing at each corpus's
drawn end ($1.4\times10^3$--$3.2\times10^4$ cases). (c)~Loop rate is
flat in subsample count (the p50 curves stay level over the 12 sizes);
separately, the BPIC'12 and BPIC'17 curves draw nearly
coincident at 59.2/59.3 loop events per 100 logged events
~\citep{vandongen2012bpi,vandongen2017bpi}. Dashed curves
in (a): the zero-parameter occupancy null at the same 12 sizes---the
expected distinct labels after $m$ of $n$ cases without replacement,
$\mathbb{E}[D(m)] = \sum_{l}\bigl(1-\prod_{i=0}^{m-1}\tfrac{n-c_l-i}{n-i}\bigr)$,
$c_l$ the number of cases carrying label $l$ (fit discussion in the
text). Semantic colour is figure-local: the panels here key corpora by hue
(green names BPIC'17); \Cref{fig:sweep} keys sweep classes by marker and
colour (axis / process logs / eval subset), \Cref{fig:router} and
\Cref{fig:endtoend} key method arms, \Cref{fig:convergence} key quantity
series; repeated hues carry no cross-figure meaning. Panel (c)'s y-axis
label echoes the left (loop-event) axis of \mbox{\Cref{fig:convergence}b} at a
different scale (compare slopes, not heights).}
\label{fig:pubsat}
\end{figure}
\begin{table}[t]
\centering
{\emergencystretch=3em\relax\raggedright
\caption{Public corpus census and fold totals, produced by the same
\texttt{fold()} routine as the SOC corpus's own folding.
Loop mass (\%) is $100(\mathrm{retry}+\mathrm{back})/\mathrm{events}$; steps
= logged events on these streams (code: \texttt{fold\_rules.py}).}\label{tab:pub}}
\vspace{1mm}
{\footnotesize
\begin{tabular}{lrrrrrrr}\toprule
corpus & cases & events & labels & advance & retry & back & loop mass (\%) \\\midrule
BPIC'12 & 13,087 & 262,200 & 24 & 107,006 & 105,776 & 49,418 & 59.2 \\
BPIC'17 & 31,509 & 1,202,267 & 26 & 489,408 & 448,205 & 264,654 & 59.3 \\
helpdesk & 4,580 & 21,348 & 14 & 19,443 & 944 & 961 & 8.9 \\
receipt & 1,434 & 8,577 & 27 & 8,332 & 6 & 239 & 2.9 \\
roadtraffic & 10,002 & 37,255 & 11 & 36,732 & 270 & 253 & 1.4 \\
statechart & 1 & 17,912 & 106 & 106 & 17,675 & 131 & 99.4 \\
\bottomrule\end{tabular}
}
\end{table}
Does the loop-collapsed action-graph picture survive outside security operations?
Following the process-mining methodology of \citet{vanderaalst2016processmining}, we
run the identical fold over six public
process-mining and software-engineering event logs (saturation curves over the
five that support subsampling)---loan applications
(BPIC'12 and BPIC'17 from the BPI
Challenge series~\citealp{vandongen2012bpi,vandongen2017bpi}, each log's fetch host and sha256 pinned in App.~B),
permit applications and service desks (receipt~\citealp{buijs2014receipt},
helpdesk~\citealp{polato2017helpdesk}),
traffic-fine handling (roadtraffic~\citealp{deleoni2015roadtraffic}) and one instrumented software-tool
session (statechart~\citealp{leemans2018statechart}): 60{,}613 cases
and 1{,}549{,}559 events in total
(\Cref{tab:pub}; URLs, sha256 provenance and a builder-correction ledger
in App.~B). Two findings generalize (i--ii); the third (iii) probes the
catalog-size boundary condition. (i)~Loop mass is corpus-generic and often \emph{larger} than the SOC
corpus's 29.2\%: 59.2\% on BPIC'12 (immediate repeats alone are 105{,}776
of 262{,}200 events) and 59.3\% on BPIC'17 (448{,}205 retries and
264{,}654 back returns inside 1{,}202{,}267 events), against 8.9\%
(helpdesk), 2.9\% (receipt) and 1.4\% (roadtraffic). The instrumented
software session is the extreme case: 17{,}675 of its 17{,}912 events
are immediate repeats and another 131 are back returns (99.4\% loop
mass counting both event kinds)---the fold compresses the
session's 17{,}912-event stream into 237 kept frames (106 first-visit
advances plus 131 returns).
(ii)~Graph growth saturates decisively in nodes (\Cref{fig:pubsat}):
distinct-node discovery flattens near the catalog ceiling wherever
the vocabulary is small, while forward-pair ($E^{+}$) discovery keeps a long tail
to the case boundary---the E1 node-vocabulary convergence is not
SOC-specific. The dashed occupancy null (\Cref{fig:pubsat} caption) matches every
observed median to a mean absolute deviation of 0.2--1.0 nodes on the
five process logs (0.2--1.2 nodes with the SOC stream, \mbox{\Cref{fig:convergence}}, as a sixth, undrawn point: 1.2 of its 154
nodes, 0.78\% of that catalog; normalized deviations run 0.78\%--4.06\%
across the six streams): for node discovery,
conditional on the observed label supports, saturation is combinatorics
rather than a lucky property of one deployment.
(iii)~The same held-out decision construction (recipe unchanged;
option and edge vocabularies rebuilt per corpus from its training fold)
carries to open tool-use trajectories: 26{,}999 ToolBench multi-step
chains~\citep{qin2023toolllm} over a 7{,}318-name API catalog ($47.5\times$
the SOC's 154 names), the large-catalog regime where the edge-ID
variant (\S\ref{sec:limitations}) is hypothesized to pay. API-Bank---a 50-name
nested-call set at the other end of the catalog scale---adds 261 loaded
dialogues from 263 scenario files~\citep{li2023apibank} (its 335-row
training fold fell below the catalog-size sweep's training floor). The
sweep results for the ToolBench catalog are summarized in App.~B.

Beyond these corpus findings comes the readout-level decomposition. The
majority arm
claims 26.8\% on the SOC stream but jumps to 66.3\% on BPIC'17 and to
87.2\% on helpdesk, where that stream's feature MLP (87.3\% masked) sits inside
half a point of its own majority baseline---once the majority arm itself runs this high, the readout has little left to add. Margins over the majority arm vary
sharply by stream---+6.0 points on SOC (the feature-MLP CE arm: 32.8\% masked against 26.8\%),
+22.7 on BPIC'17 (the feature-MLP CE arm: 89.0\% masked against 66.3\%),
+26.4 on BPIC'12 (the feature-MLP CE$+$Brier arm: 80.8\% masked against 54.4\%)---reported
here descriptively. A variable-order Markov baseline with
backoff-style interpolation (the backoff Markov; the standard prefix-scoring
baseline of predictive process monitoring; App.~B) probes how much
of each margin is memorizable transition structure: contexts over the
traced node flow (the last 0--3 nodes of history; order counts context
nodes, so the order-3 table conditions on a 3-node history---a 4-gram
counting its target) are
counted, orders interpolated to the marginal, and the table
option-gated like the deployed masked-renorm gate. On the five process logs (BPIC'12, BPIC'17, helpdesk,
receipt, roadtraffic) the order-3 table (each log's per-stream best) collects most of each
margin (within $\pm 2.5$ points): 89.7\% on BPIC'17 against the feature
MLP's 89.0\%, and 83.9\% on receipt against the feature MLP's 86.4\% (CE$+$Brier; its majority arm 75.1\%). On the two large agent
streams---SOC and the 7{,}318-name ToolBench training
catalog---transition tables fall short of the readout: the best
order-$k$ table stops at 31.1\%, short of the feature MLP's 32.8\% on
SOC, and at 7{,}318 names no order-$k$ table even reaches the 37.6\%
majority arm that the CE$+$Brier arm beats by 3.2 points. The featurized residual (MLP minus
the table's per-stream best) is within $\pm 2.5$ points across the five
process logs and positive at the point estimate on both large agent
streams---a comparison whose caveats run to the table's benefit (order
chosen as its per-stream best on the same test split; moving the choice to the
validation fold, which every public stream already carries, leaves the picture
unchanged: validation picks the same order on six of the nine streams, and on
the other three it picks one whose test accuracy is at or below the
test-selected order's (largest gap 0.6 points, ToolBench evaluation), so the
validation-selected column never flatters the table; the public folds are
already session-isolated---0 of 17{,}942 test traces appear in training, and a
trace-disjoint refit reproduces every order's accuracy exactly; point estimates,
no intervals). The residual, rather than the raw margin over the majority
arm, is what the feature MLP adds beyond order-$\leq$3 backoff
transition tables---a residual that is conservative under the on-test
order selection above---not beyond stream statistics at large (the
feature MLP's own inputs are stream statistics).
ToolBench-eval (the 351-name evaluation subset, 161 hops) and API-Bank sit at the hard end and motivate
the cautions below. There the ToolBench runs reproduce
the SOC's masked-below-declared direction, with a gap far wider than
the SOC's own (32.8\% masked versus 35.6\% within the declared space, the CE
arm): 34.9\% masked against 66.9\% within
the declared option space (CE$+$Brier arm)
on the 351-name evaluation subset
(47.8\% of test targets leave the declared options), and 40.8\% against
48.2\% on the 7{,}318-name training catalog (CE$+$Brier; 15.4\% out-of-space).
Four cautions accompany the comparison. (i)~The majority arm (37.6\%) edges
the multinomial logistic-regression arm (35.7\% masked, 36.4\% raw) at 7{,}318 classes (an
800-iteration budget under-converges a linear arm at this scale).
(ii)~The small nested-call set reverses the logistic/MLP order (API-Bank:
logistic 61.5\% versus the feature MLP's 56.9\% on the CE$+$Brier arm; the stream's interpolated
tables reach 59.6\%---above the feature MLP and the majority arm but
below the logistic arm).
(iii)~On the same set the edge-ID readout leads the class-space readout
(59.0\% versus 55.6\% masked, both CE arms).
(iv)~On the 351-name
evaluation subset (161 held-out hops) the order-2 table (a 2-node
history---the 3-gram) leads
(36.0\% masked/69.0\% in-space versus the class-space readout's
34.9\%/66.9\%). The controlled catalog-size sweep in
App.~B tracks which of these effects moves with catalog size (narrower
budgets also carry narrower train/test splits here).

\paragraph{E9: what the induced graph is worth (necessity ablation).}
The probe arm reads only the frozen encoder state, so on that arm the induced graph
enters solely through the declared option space $A(v)$. Holding features, readout
(linear probe), recipe, seeds and budget fixed and changing only the evaluation-time
option space, the declared space reaches 30.67\% while a flat 155-class catalog
reaches 27.07\% ($-3.5$ points), a size-matched global train-frequency space 29.60\%
($-1.4$), and a random $K$-set 8.4\% (\Cref{tab:e9}); a re-implementation of the
published probe reproduces the frozen headline to four decimals (30.67\% mean masked
agreement against 30.67\%; per-seed spread 0.0037 against 0.0039), so the arms are
comparable. A node-conditioned space that uses the same node identity but no graph
traversal reaches 30.79\% ($+0.1$ points; paired trace-clustered bootstrap CI
$[0.0, +0.3]$ over 2{,}000 resamples; 6 wins, 1 loss, 3{,}868 ties), and the union of
targets ever observed at that node gives the same number. The additivity is the
reading we claim: flat $\to$ size-matched frequency prior $+2.5$ points, frequency
prior $\to$ node conditioning $+1.2$, node conditioning $\to$ graph traversal
$\approx 0$. Two limits on that reading: the arm is the probe, not the feature MLP
(where the graph additionally enters through engineered node features), and the
node-conditioned competitor uses the targets observed at the node, which is close to
the graph's own outgoing-edge set (mean 40.4 vs.\ 40.2 options); what the ablation
separates is therefore node identity from graph traversal, not node information from
none. Artifacts are listed in the Reproducibility section.
\begin{table}[t]
\centering
\small
\caption{Graph-necessity ablation (E9/E9b). Only the evaluation-time option space
changes; training, features, readout recipe and budget are identical across arms
(5 seeds). $\Delta$ and CIs are paired against \texttt{declared} with a
trace-clustered bootstrap (2{,}000 resamples, seed-0 model); ``---'' is the
reference arm.}
\label{tab:e9}
\begin{tabular}{lrrr}
\toprule
option space at evaluation & mean $|A|$ & agreement (\%) & $\Delta$ vs declared (95\% CI) \\
\midrule
declared $A(v)$ (induced graph) & 40.2 & \textbf{30.67} & --- \\
node-conditioned, no traversal & 40.4 & 30.79 & $+0.13$ $[0.00, +0.27]$ \\
global train-frequency, size-matched & 40.2 & 29.60 & $-1.37$ $[-1.88, -0.91]$ \\
flat catalog (all 155 classes) & 155 & 27.07 & $-3.54$ $[-4.10, -2.95]$ \\
random $K{=}|A(v)|$ (floor) & 40.2 & 8.36 & $-22.45$ \\
\bottomrule
\end{tabular}
\end{table}

\section{What the typed interface can replace, and what it cannot}\label{sec:replace}

Earlier sections score one realised next hop. This section asks a narrower question with a deployment
consequence. Which part of that decision can a small trained model take over, and which part needs an
agent that can see more of the episode than the released state exposes? Every number below is read
from the held-out fold or from an arm we ran for this section, and each table names the artefact its
columns come from.

\paragraph{E23a: scale does not move the zero-shot reading.}
jpt-4b, a 4B general-purpose checkpoint asked to choose among the declared options, reaches 9.0\% agreement
(9.9\% within the declared space) on a seeded 200-decision sample. On the typed model's
twenty-candidate shortlist it reaches 9.5\%, or 10.5\% in space (\Cref{tab:e23cascade,tab:e23scale}).
The domain encoder reaches 8.6\% zero-shot on the full fold. The diagnostic is less flattering. Its
accuracy equals the sample's majority-class rate, and it picks the candidate printed first
22\% of the time against a 5\% uniform rate. Order was shuffled per decision, so the bias cannot
pass for competence. On the same 200 decisions the fine-tuned encoder reaches 31.0\%. Tuning is what separates the two families here, so \Cref{tab:e23scale} adds the tuned reading on the full fold.

\paragraph{E23b: fine-tuning moves it, and fine-tuning is what moves it.}
Freezing the encoder and training the decision head alone gives 27.7\% (30.1\% in space). Unfreezing
the encoder for three epochs gives 32.2\% (34.9\%). The 4.5-point difference is attributable. The two
arms share the renderer, the option set, the optimiser and the epoch budget, and differ only in which
parameters receive gradients. The gradient path is the whole effect. The fine-tuned arm sits 0.65
points below the strongest feature readout in \Cref{tab:router} (32.8\% and 35.6\%). The typed
interface reaches parity with the hand-built feature arm.
Three numbers of similar size appear in this paper from three protocols rather than one
quantity. 32.8\% is the strongest hand-built feature readout, 32.2\% the fine-tuned typed
encoder measured here, and 32.15\% a full-supervision run under the loop protocol of
\Cref{sec:rsi-measured}. All three score the same frozen held-out fold.

\paragraph{E23c: narrowing the option set does not substitute for state.}
The fine-tuned model ranks the realised next hop inside its top twenty for 90.7\% of the fold.
Narrowing the candidate set does not hand the decision to a general model. On a random 800-decision
subset jpt-4b scores 9.9\% over all declared options, 12.1\% over twenty candidates and
19.9\% over five, while the typed encoder reaches 32.2\% over all options on the full fold
(\Cref{tab:e23cascade,tab:e23scale}). A shortlist that contains the answer almost nine times in ten
does not let a zero-shot general model find it. Tuning the same checkpoint on the same protocol and fold closes almost all of that gap. It reaches 31.9\% against the typed encoder's 32.2\% on the full held-out fold, a difference of 0.25 points. An exact paired test over the same 3{,}875 decisions does not resolve it: 229 decisions the tuned model gets and the typed encoder misses, 239 the other way, $p=0.68$. On the shared 800 decisions its agreement rises from 9.9\% to 33.0\% ($p<10^{-3}$). What the typed encoder keeps is cost, 10\,ms against 0.24\,s per decision, and a tenfold smaller parameter count. The accuracy advantage over a language model belongs to tuning, not to the interface. What binds is the state text, not the number of candidates.

\paragraph{E23d: task intent helps by about one point; more context does not help at all.}
Three arms share one training recipe, one seed and one batch size, and differ only in the state string
(\Cref{tab:e23state}). Adding the decision's own goal field, a six-valued enum, lifts agreement by
1.08 points, from 32.00\% to 33.08\%. Replacing the eight-action window with the full action prefix,
while keeping the goal, drops it by 1.39 points. Length is not the lever. The released records carry
the goal field and the published renderer never exposes it, so this increment is available today at no
extra cost. Both increments are single-run point estimates from one harness at one seed, and the
1.08-point gain sits inside the 1.9-point cross-run spread reported in \Cref{sec:rsi-measured}.
The larger claim, that a plan or playbook in the state would lift the in-graph decision
far higher, is the interface change this paper cannot yet measure.

\paragraph{E23e: the released state carries a measurable ceiling.}
To bound what any state-conditioned predictor can reach here, we grouped held-out decisions by an
increasingly specific key (\Cref{tab:e23ceiling}). Each group kept only decisions from other traces,
and we asked how often a decision agrees with the modal choice of its group. The construction is
leave-one-out and cross-trace, so a trace never votes for itself. The reading rises with the key and
then flattens: 25.6\% for the station alone, 26.5\% with one preceding action, 38.6\% with three,
42.8\% with the full released prefix. Requiring at least two other-trace observations raises the last
figure to 44.9\%, but over a shrinking support: the group behind 25.6\% holds all 3{,}857 decisions,
while the group behind 44.9\% holds only the 1{,}234 (31.8\%) whose key recurs in another trace,
and the levels are therefore not scored over a common denominator. With everything the release
exposes, repeated states still disagree more often than
they agree. Agreement with a single realised sample cannot approach one on this fixture. No confidence
rule can lift the accepted subset to 90\% at any coverage, and no model size changes that: the
constraint is the state, not the capacity.

\begin{figure}[t]
\centering
\includegraphics[width=\linewidth]{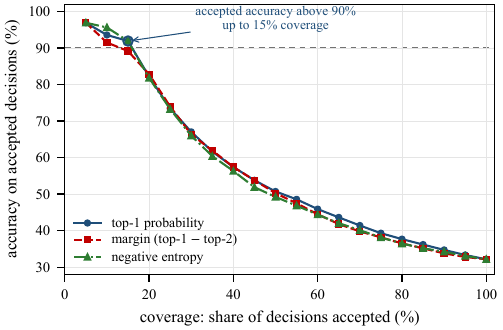}
\caption{Accept-or-escalate frontier of the fine-tuned typed encoder on the held-out fold, under three
confidence rules. Accepted accuracy stays above 90\% up to 15\% of coverage, then falls towards the
accuracy of the whole fold as coverage approaches one. The three rules trace nearly the same curve, so
the frontier reflects the state rather than the choice of score.}
\label{fig:e23frontier}
\end{figure}

\paragraph{E23f: two questions, two operating points.}
Containment and agreement must not be conflated. For containment the shortlist reaches 90.7\% at
twenty candidates. The residual 9.3\% needs an agent holding the full episode. The share of held-out
targets that fall outside the declared options altogether is 7.9\%. Both figures sit under the 20\%
escalation budget a deployed cascade can afford. For agreement, we sorted decisions by top-1
probability and kept the most confident subset (\Cref{tab:e23frontier}, Figure~\ref{fig:e23frontier}).
Accepted accuracy holds above 90\% up to 15\% coverage, 91.9\% at 15\%, 82.2\% at 20\%, 37.7\% at 80\%.
A confidence gate therefore buys a small high-precision band. The band is not an artefact of the score:
at 15\% coverage, margin gives 89.2\% and negative entropy 91.9\%. The trained encoder answers the
containment question; the ceiling above bounds the agreement question.

\paragraph{E23g: agreement tracks task outcome, modestly.}
Readouts in \Cref{tab:router} are scored against the fleet's realised choices. We also asked what
those choices predict. Over the 345 held-out traces, a median split on agreement gives 68.8\%
completion in the upper half against 62.2\% in the lower half. Completed traces are longer, at 15.5
steps against 10.3, so we repeated the split inside step-count strata. The gap sits in mid-length
traces: +46.0 points at 6--10 steps, separating 22.2\% completion from 68.3\% over 126 traces,
against +2.0 at 2--5, +12.5 at 11--20 and +22.4 at 21 or more, where both halves clear 70\%. This is
an association. A mechanism would need a controlled rollout, and the labels are model-derived.

\begin{table}[t]
\centering
\small
\caption{What the typed interface can take over, on the held-out test fold. The two
zero-shot language-model rows were drawn on a seeded 200-decision sample of the same fold and
are accompanied by the typed encoder measured on that same sample, so that comparison is
same-denominator; the LoRA-tuned jpt-4b row is scored on the full fold instead. ``cands'' is the number
of options offered to the deciding model. The last row is the containment ceiling of the shortlist,
not the accuracy of any model.}
\label{tab:e23cascade}
\vspace{1mm}
{\footnotesize
\resizebox{\linewidth}{!}{
{\setlength{\tabcolsep}{4.5pt}\begin{tabular}{lrrrrl}\toprule
decision rule & acc.\ (\%) & in-space (\%) & cands & $n$ & note \\\midrule
majority arm & 26.8 & -- & 1 & 3{,}875 & relevance floor \\
feature MLP, CE$^{\dagger}$ & 32.8 & 35.6 & 1 & 3{,}875 & strongest prior readout \\
typed encoder, frozen & 27.7 & 30.1 & 1 & 3{,}875 & parity arm \\
typed encoder, fine-tuned & 32.2 & 34.9 & 1 & 3{,}875 & 3 epochs, seed 0 \\
typed encoder, fine-tuned & 31.0 & -- & 1 & 200 & same sample as the rows below \\
jpt-4b, all options & 9.0 & 9.9 & 46 & 200 & zero-shot \\
jpt-4b, top-20 shortlist & 9.5 & 10.5 & 20 & 200 & zero-shot \\
jpt-4b, LoRA-tuned & 31.9 & 34.6 & 40 & 3{,}875 & fine-tuned, full fold \\
top-20 shortlist oracle & 84.0 & 90.7 & 20 & 200 & 90.7\% on the full fold \\
\bottomrule\end{tabular}}
}}
\end{table}

\begin{table}[t]
\centering
\small
\caption{State conditioning. The three arms share one training recipe, one seed and one batch size and
differ only in the state string handed to the typed encoder; the published renderer is the ``last eight
actions'' arm. The goal field is present in the released records and absent from the published
renderer. ``top-20'' is masked recall over each state's declared options.}
\label{tab:e23state}
\vspace{1mm}
{\footnotesize
\resizebox{\linewidth}{!}{
{\setlength{\tabcolsep}{4.5pt}\begin{tabular}{llrrr}\toprule
state given to the typed encoder & adds & acc.\ (\%) & in-space (\%) & top-20 (\%) \\\midrule
last eight actions & published renderer & 32.00 & 34.75 & 89.74 \\
last eight actions $+$ goal & task intent & \textbf{33.08} & \textbf{35.93} & 89.24 \\
goal $+$ full own history & length, not intent & 31.69 & 34.42 & 89.18 \\
\bottomrule\end{tabular}}
}}
\end{table}

\begin{table}[t]
\centering
\small
\caption{The information ceiling of the released state. ``covered'' counts held-out decisions whose
key also occurs in another trace; ``agree'' is the leave-one-out modal agreement among those other
traces; the last column requires at least two such other-trace observations. The key is made more
specific from top to bottom. L4 and L5 differ in kind: L5 adds engineered features instead of history
and does not improve on L4.}
\label{tab:e23ceiling}
\vspace{1mm}
{\footnotesize
\resizebox{\linewidth}{!}{
{\setlength{\tabcolsep}{4.5pt}\begin{tabular}{llrrr}\toprule
key & state content & covered & agree.\ (\%) & agree., $\ge 2$ (\%) \\\midrule
L1 & current station & 3{,}857 & 25.6 & 25.6 \\
L2 & one preceding action & 3{,}400 & 26.5 & 27.6 \\
L3 & three preceding actions & 1{,}694 & 38.6 & 43.4 \\
L4 & the full action prefix & 1{,}234 & 42.8 & 44.9 \\
L5 & L3 plus observable features & 1{,}301 & 41.2 & 44.7 \\
\bottomrule\end{tabular}}
}}
\end{table}

\begin{table}[t]
\centering
\small
\caption{The accept-or-escalate frontier of the fine-tuned typed encoder on the held-out fold, sorted by
top-1 probability. Keeping the most confident 15\% of decisions holds accepted accuracy at 91.9\%;
the curve has no 90\% point beyond 15\% coverage. Temperature was fitted on the validation fold
(fitted value 1.0). AURC is 0.426.}
\label{tab:e23frontier}
\vspace{1mm}
{\footnotesize
\resizebox{\linewidth}{!}{
{\setlength{\tabcolsep}{4.5pt}\begin{tabular}{lrrrrrrrr}\toprule
accepted coverage (\%) & 5 & 10 & 15 & 20 & 30 & 50 & 80 & 100 \\\midrule
accepted accuracy (\%) & 96.9 & 93.6 & 91.9 & 82.2 & 67.0 & 50.7 & 37.7 & 32.2 \\
\bottomrule\end{tabular}}
}}
\end{table}

\begin{table}[t]
\centering
\small
\caption{Same-denominator comparison on the identical held-out question: choose the realised next hop
among the declared options. Every jpt-4b row but the last is zero-shot; the last is
LoRA-tuned on the training fold, which makes it the like-for-like opponent for the typed encoder.
The typed encoder is fine-tuned. Latency is one
decoder pass for the language model and one forward pass for the typed encoder.}
\label{tab:e23scale}
\vspace{1mm}
{\footnotesize
\resizebox{\linewidth}{!}{
{\setlength{\tabcolsep}{4.5pt}\begin{tabular}{lrrrl}\toprule
decision rule & params & acc.\ (\%) & per decision & note \\\midrule
jpt-4b, all options & 4B & 9.9 & 0.22\,s & zero-shot, decoder pass \\
jpt-4b, LoRA-tuned & 4B & 31.9 & 0.24\,s & fine-tuned, full fold \\
jpt-4b, 20 candidates & 4B & 12.1 & 0.14\,s & zero-shot, shortlist \\
jpt-4b, 10 candidates & 4B & 14.4 & 0.11\,s & zero-shot, shortlist \\
jpt-4b, 5 candidates & 4B & 19.9 & 0.10\,s & zero-shot, shortlist \\
Laya zero-shot typed choice & 421M & 8.6 & 10\,ms & domain encoder, no tuning \\
typed encoder, fine-tuned & 421M & 32.2 & 10\,ms & this work \\
\bottomrule\end{tabular}}
}}
\end{table}

\section{What a gated loop adds over full supervision}\label{sec:rsi-measured}

\Cref{sec:replace} bounded a single decision. The deployment question is whether a loop can raise that
bound on its own, by turning the graph's own unlabelled traffic into training rows. This section runs
the loop and reports what it does, including where it fails.

\paragraph{E25a: protocol.}
We split the training fold by trace into an anchor set of 3{,}071 decisions and an unlabelled pool of
9{,}542, and froze the test fold. Each round draws 600 pool rows, admits some of them by the arm's own
rule, retrains the typed encoder for two epochs on the buffer and re-scores the frozen fold. The gate
sees no label for the row it admits. It compares the current model's argmax against a variable-order
backoff table built only from buffer rows and admits the row when the two agree. A random arm admits
a fixed number of rows per round, and a third arm runs the same accounting with no typed model at all,
using that variable-order backoff table itself as the predictor, a Markov-style control
(Table~\ref{tab:e25rsi}).

\begin{figure}[t]
\centering
\includegraphics[width=\linewidth]{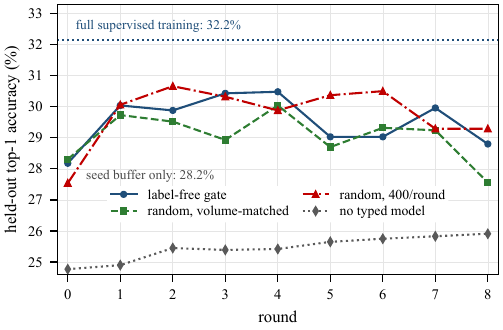}
\caption{Held-out top-1 accuracy of the typed encoder over eight rounds of gated self-training, for four
acquisition rules. The three typed arms plateau between 29\% and 31\% and end below their own peak;
the model-free arm climbs steadily from a lower floor and never reaches them. The dashed line is full
supervised training on the same fold, at 32.15\%, which no arm reaches. Round 0 is each arm's own
anchor set, so the starting points differ slightly (seed 0; the round-0 spread over seven runs
is 1.9 points).}
\label{fig:e25loop}
\end{figure}

\paragraph{E25b: the loop moves, and it stops below supervised training.}
Every typed arm ends above its own anchor-only starting point. The peaks come before the last round:
at seed 0 the gate peaks at 30.48\% in round 4 and ends at 28.80\%, averaging 29.39\% over three
seeds. The induced graph grows from 1{,}015 to
1{,}257 edges (Figure~\ref{fig:e25loop}). Full supervised training on the same fold reaches 32.15\%.
No arm closes that gap. The high-precision band also stays at half its width: accepted accuracy holds
above 90\% up to 6.8\% of coverage under the gate, against 15\% for the fully trained head. The loop
is a real but limited mechanism here. What it cannot do is manufacture the information the state
lacks.

\paragraph{E25c: controls.}
Two controls bound the reading. First, selection. Each arm draws its own anchor set, so round-0 scores
spread over 1.9 points (26.45\% to 28.31\%) across the seven runs we have; seed 0 alone gave 0.8
points. That spread is our noise floor. At matched admitted volume the gate averages 29.39\% over
three seeds (28.80, 29.37, 29.99). Random admission averages 28.77\% (27.56, 29.45, 29.29). The
per-seed gaps are $+1.24$, $-0.08$ and $+0.70$ points, a mean of $+0.62$. A sign test over the three
paired seeds gives two positives and one negative ($p=1.0$). The gap therefore sits inside the floor, and at this scale the gate's
advantage over volume-matched random admission is inside seed noise. We report the gap as a point
estimate with its seed range. The unrestricted random arm admits 43\% more
rows and lands 0.49 points above the gate at seed 0 (one run). More unselected data thus roughly
compensates for worse selection. Selection quality is a weak effect at this scale. Acquisition volume
is the stronger term. Second, capacity. With no typed model the same
accounting reaches 25.91\%, which is 2 to 3 points below the typed arms. Its 90\%-accuracy band
collapses to 0.001. The typed encoder is what makes an accept-or-escalate band exist at all.

\paragraph{E25d: what grows, and what that does not show.}
Edge counts rise in every arm, but they track how much of the pool an arm consumed: the model-free arm
consumes the most and finds the most edges. Growth of the graph is therefore evidence that the loop
explores, not that the gate selects well. Combining the two sections, the honest summary of the
deployment claim is this. The typed interface replaces the per-decision deliberation with a
millisecond forward pass and reaches parity with the strongest feature readout, containment is 90.7\%
at twenty candidates, and escalation needs to cover under 10\% of decisions. Raising in-graph accuracy
past a third of the fold is not a matter of more self-training, more parameters or more deliberation.
It needs a state that carries the plan and the observation, which the released records do not.

\paragraph{E25e: what a deployment should do with this.}
Three operating points follow from the two sections, and they are meant to be read
together. Inside the graph, route with the typed encoder and accept its decision: one forward
pass, no generated text, 19.5\,ms per decision in the typed-choice form and 10\,ms for the
encoder pass the replay uses. At the top of the confidence ranking, accept a band rather
than the whole fold: 15\% coverage holds 91.9\% agreement. Return the rest, because the
realised hop lies outside the declared options for 7.9\% of decisions and outside the model's
top twenty for 9.3\%, so a budget under 20\% covers both measures. Returning a decision is
escalation, and its price is measured: on the replay stream the accept/escalate hybrids agree
with the reference on 20.9\% / 19.1\% / 16.5\% of decisions at escalation rates of
74.1\% / 80.2\% / 85.0\% (\Cref{tab:endtoend}), below the pure readout's 30.7\% at zero calls,
and the escalated remainder costs 2.2\,s a decision, so an 85\% budget moves mean decision time
from 2.2\,s to about 1.9\,s. The band is a latency and load instrument rather than an agreement
gain, and it assumes an escalation arm stronger than the zero-shot stand-in these hybrids were
measured against. Above that budget none of
the remedies tested here applies: more self-training, more parameters and more deliberation
each leave the ceiling where the state puts it. The escalation rule itself must carry
per-state competence rather than confidence alone, since the confidence-gated cascade lost
agreement at every threshold we tested.

\begin{table}[t]
\centering
\small
\caption{The gated self-training loop over eight rounds, held-out test fold. ``admitted'' counts the
pool rows the arm added across all rounds, so the arms differ in both rule and volume; the third row
is the volume-matched random control for the first. ``edges'' is the number of distinct forward
transitions in the induced graph at round 8. ``90\% cov.'' is the largest fraction of decisions that
can be accepted while holding accepted accuracy above 90\%. The last two rows are not loop arms: they
bracket it from below and above. The four loop rows are seed 0; three-seed means and the seed
spread are given in the text.}
\label{tab:e25rsi}
\vspace{1mm}
{\footnotesize
\resizebox{\linewidth}{!}{
{\setlength{\tabcolsep}{4.5pt}\begin{tabular}{lrrrrrr}\toprule
acquisition arm & admitted & edges & acc.\ (\%) & in-space (\%) & 90\% cov. & @20\% cov. \\\midrule
label-free gate & 2{,}283 & 1{,}257 & 28.80 & 31.28 & 0.068 & 71.6 \\
random, same cap (400/round) & 3{,}200 & 1{,}543 & 29.29 & 31.81 & 0.070 & 73.3 \\
random, volume-matched (280/round) & 2{,}240 & 1{,}395 & 27.56 & 29.93 & 0.070 & 72.8 \\
no typed model (frequency backoff) & 4{,}800 & 1{,}741 & 25.91 & 28.14 & 0.001 & 63.5 \\
\midrule
seed buffer only, no rounds & 0 & 1{,}015 & 28.18 & 30.61 & 0.153 & 74.8 \\
full supervised train fold & 12{,}613 & --- & \textbf{32.15} & \textbf{34.92} & 0.150 & --- \\
\bottomrule\end{tabular}}
}}
\end{table}

\section{Limitations and Discussion}
\label{sec:limitations}
\textbf{Replay evaluation.} The held-out logs are a realized environment,
not a live one: policies are scored against the deployed fleet's own
choices (teacher-forced scoring, strict survival and per-wrong-hop gap
costs), so a
policy that would \emph{improve} on the reference cannot demonstrate that
improvement here---agreement with the fleet at a fraction of its decision cost is the claim.
Closed-loop rollouts with tool results reattached (this needs a fuller
export than the stripped one described here) are the natural next step. The
graph-free comparison that scores every catalog action as one flat option
space is run here (E9, Reproducibility): the flat catalog loses 3.5 points and a
size-matched global-frequency space loses 1.4 points, so restricting candidates per node
is load-bearing, but a node-conditioned space built from per-node train statistics
without graph traversal matches the graph-declared space within noise ($+0.1$ points,
95\% trace-clustered CI $[0.0, +0.3]$). The induced graph's measured value here is
therefore as a construction that yields node-conditioned option sets, not as a
traversal whose specificity this data can resolve. \textbf{Derived labels.} Task outcomes
come from the close-out convention (\S\ref{sec:setup}): the convention
is machine-checkable and checked against an assistant's manual
session-log re-reading (30/30), but it certifies
\emph{declared} completion,
not the semantic quality of the report text---which the export strips.
The unconstrained-judge failure mode we quantify (38.5\% agreement when
the judge expects report text in a field that never carries it) is a
caution for LLM-judge pipelines generally: calibrate against a manual
session-log re-reading, not against another judge. \textbf{Scope of the claims.} We evaluate a decision
layer over logs that already exist. Authorization, retention and
re-identification boundaries of the underlying sessions, the human review roles
around them, change management for catalog updates, and the explicability a
planner needs when it cannot verify its own outputs are properties of the
deployment that
produced the logs; we do not measure them, and our audit statements are limited
to what the export can show (which, as noted above, includes no separate human
auditor in the loop). \textbf{State reconstruction.} Tool \emph{result} bodies and
chat bodies are absent from the export (the tool-result field is null in all
22{,}192 tool records; call arguments survive, below), so the router's context is
structural (labels, ordering). Two result-adjacent families do survive in
the export and go unused in the main pipeline---per-call
latency (100\% of tool calls) and the call arguments (98.9\%)---so
their contribution is already measurable without the payloads: an artifact-shape-verified reproduction of the \S\ref{sec:results} feature-MLP CE$+$Brier arm scores 32.7\% masked agreement, and appending prefix-latency / latency-plus-argument-shape features adds +0.4 / +1.3 points (five seeds, \texttt{feat\_lat\_results.json}, Reproducibility)---a small but real measured gain that stops this limitation's prediction being an untested assertion.
\textbf{Domains and slices.} One security-operations deployment, four
stage slices; the protection-stage zero close-out rate (App.~A) means
slice comparisons conflate routing behavior with pipeline role.
\textbf{Edge IDs.} The per-edge identifiers do \emph{not}
beat the class-space readout at the SOC catalog size with our shallow state
features---the small API-Bank set and the 351-name ToolBench-eval subset are
the material small-catalog reversals
(\S\ref{sec:results}, \Cref{fig:sweep})---and the natural counter---that verdict being a
small-catalog artifact---is now tested: a controlled catalog-size sweep
on the 7{,}318-name ToolBench catalog (\Cref{fig:sweep},
App.~B) holds recipe and features
fixed while shrinking the option vocabulary to the top-$C$ actions by train frequency ($C$, the kept catalog budget). The
row-filtered sweep gives a class advantage of net +0.9 points at $C{=}154$
to +1.3 at $C{=}6{,}750$ (a dip to +0.6 at $C{=}512$). That sweep keeps only rows whose current action and target both lie inside top-$C$, so the scored subset itself shrinks
with $C$---by up to 65.5\% of rows at the narrow budgets. Re-running it on a
\emph{fixed} row set (out-of-catalog goals masked to an always-offered
unknown class) leaves the shape intact and
the gap smaller: net +0.5 points at $C{=}154$ to +1.2 at
$C{=}6{,}750$ (a dip to +0.3 at $C{=}512$) over \emph{strictly paired}
candidate lists---identical options for both arms on rows whose target lies on
that list; charging each arm for the options it cannot name instead flips the
sign at the two narrow budgets ($-1.8$, $-1.2$). The class-space advantage
therefore ends wider across the sweep (non-monotonically) and, in the row-filtered and paired means,
never crosses over on the measured axis (top-$C$ up to 6{,}750; outside the noise-level ties, the
charging-convention flips above and the contrary small-catalog slices---the evaluation subset at its
275-name training catalog and API-Bank---are the exceptions disclosed in \Cref{fig:sweep}): even at the
widest tested budget ($C{=}6{,}750$, $43.8\times$ the
SOC's 154 option names), a shared
class vocabulary is the better parameterization of the next-hop choice.
The edge-ID variant may still pay where option names carry no
meaning~\citep{sun2026typesafe}---that regime remains open.
\textbf{Scope.} The readout has no mechanism for multi-step internal
search (the single-pass design of~\citet{cheng2026thisthat} likewise cannot carry intermediate results through); long detours are delegated to the graph
and to LLM escalation at genuine unknowns. All typed-model timing is batch-1
GPU wall clock (NVIDIA H20, App.~B) except the in-vivo zero-shot
pipeline, measured on CPU; service and judge latencies are measured in
vivo under production load and vary with it.

\textbf{Data availability and governance.} The logs behind
\S\ref{sec:results} are production security-operations records of a deployment
that is not ours to release, and no human-subjects data were collected for this
study. The derived, non-identifying artifacts that back every reported
number---frozen per-decision predictions, the aggregation and audit scripts, and
the per-experiment receipts---can be made available on request. The public bundle is sufficient
to re-derive every corpus-level aggregate reported here---episode, session,
split and per-slice counts, the held-out option-space mean and its
$3{,}568$-of-$3{,}875$ coverage, the catalog size and the fold-event
totals---while the pre-fold trace count, the semantic families and the
model-level results require the internal export.
Authorization, retention and re-identification boundaries, the human review
roles around the sessions, and catalog change management remain properties of
the deployment that produced the logs: what this paper states about them is
limited to what the stripped export can show, and nothing here should be read as
a governance assessment of that deployment.

\section{Conclusion}
Logs already contain an organization's workflow knowledge; \method{} turns
them into a graph whose transition set runs a long tail and whose node
vocabulary converges at rates a zero-parameter occupancy null already
predicts (E1), and pairs it with a typed router that reads
next-hop decisions off that graph in one forward pass, escalating only
its own uncertainty to the slow, expensive path. The loop closes: every executed
walk becomes gated training signal. On 1{,}615 production episodes the
pieces separately hold---quantified node-vocabulary convergence of the induced action
graph (E1); feature-MLP readouts reaching 32.8\% agreement with the deployed
fleet (35.6\% within the declared space), with the hidden-state probe
variant holding 30.7\% at a 10\,ms per-state encoder pass plus a linear
head against
2.2\,s LLM hops, and out-agreeing the re-prompted MIMO Pro V2.6 stand-in
(tool-result-stripped, zero-shot) on
the same states (E2--E3)---a transfer-floor comparison, not a capability
ceiling: the stand-in is never fine-tuned. Route-level replay keeps the typed layer in the
agreement/cost corner while exposing the accept/escalate hybrid's
calibration failure as a negative result (E3). A self-evolution loop's
rollback gate scores every candidate while the loop absorbs the
flipped labels of one update batch (12\% of that batch's labels corrupted---about
0.3\% of the 12{,}613-decision base) without firing rollback on the plotted seed-0 chain (E4). Across ten seeded chains the gate refuses 13 of the 40 candidate updates, in 6 of the 10 chains. An external replication runs the
same fold over 60{,}613 public cases, where loop mass reaches 99.4\% on an instrumented software session and
1.4\% on traffic fines; node
discovery flattens wherever catalogs are small (E5); \method{}'s decision
construction carries over to a 7{,}318-name tool catalog, where a
transition table isolates what the feature MLP contributes beyond
order-$\leq$3 memorized transitions; on the case streams an occupancy
null shows node saturation is combinatorics, conditional on the
observed label supports. The
invitation is concrete: drop your logs into the fold pipeline and measure
your own saturation curve.

\section*{Reproducibility statement}
{\raggedright
All model experiments run on the released pipeline in
\texttt{experiments/} inside an
\texttt{nvcr.io/nvidia/pytorch:26.03-py3} container on a remote host with
$2\times$ NVIDIA H20 (GPUs 0/1) against frozen artifacts in
\texttt{experiments/real/} (\texttt{episodes.jsonl},
\texttt{e2\_*.jsonl},
\texttt{e1\_saturation.json}, \texttt{e2\_results.json},
\texttt{e2\_layaprobe\_results.json}, \texttt{e2\_laya\_results.json},
\texttt{gpu\_latency.json}, \texttt{e3\_replay.json},
\texttt{e4\_rsi.json}, \texttt{e8\_slices.json},
\texttt{outcome\_labels\_v2.jsonl}, \texttt{outcome\_labels.jsonl}, \texttt{corpus\_census.json},
\texttt{e2\_laya\_raw.json}, \texttt{pub\_e1\_results.json},
\texttt{pub\_sweep\_results.json}, \texttt{pub\_coupon.json},
\texttt{pub\_markov\_results.json}, \texttt{e2\_meta.json},
\texttt{e2\_results\_cpu\_local.json}, \texttt{forensics\_stats.json},
\texttt{gold\_30.json}, \texttt{audit\_sample.json},
\texttt{outcome\_label\_summary.json}, \texttt{outcome\_label\_v2\_summary.json},
\texttt{tool\_parse\_report.json}, \texttt{tool\_manifest.json},
\texttt{pub\_fetch\_manifest.json}, \texttt{e3\_mimo\_stream.json} and
every public stream's \texttt{*\_e2\_results.json}/\texttt{*\_e2\_meta.json} pair; the typed-choice feature tensor
\texttt{feats\_laya\_cls.npz}, assembled from the sharded extraction passes by \texttt{assemble\_feats.py}, is a
further frozen input). The eight stream names are
\texttt{apibank}, \texttt{bpi12}, \texttt{bpi17}, \texttt{helpdesk},
\texttt{receipt}, \texttt{roadtraffic}, \texttt{toolbench\_train} and
\texttt{toolbench\_eval}. The \texttt{experiments/addendum/} artifacts
(\texttt{feat\_lat\_results.json}, \texttt{e4\_rsi\_multi.json},
\texttt{pub\_sweep\_fixed\_results.json}, \texttt{e3\_replay\_multi.json},
\texttt{e9\_graph\_necessity.json}, \texttt{e9b\_node\_prior\_curkey.json}, \texttt{e9b\_node\_prior\_nodekey.json}) are
likewise frozen inputs (the E9/E9b option-space arms are produced by
\texttt{e9\_graph\_necessity.py} and \texttt{e9b\_node\_prior.py}, which read only the
frozen feature tensor, the derived decision streams and the frozen probe reference,
and which reconcile to \texttt{e2\_layaprobe\_results.json} as a parity check), and \texttt{tables/*.tex} and figure PDFs are regenerated from these
artifacts by \texttt{make\_tab\_corpus.py},
\texttt{render\_figs\_tabs\_real.py} and \texttt{render\_pub.py}.\par} The raw corpus export is internal
SOC telemetry and cannot be redistributed; the \emph{derived} corpus joins
the code release in pseudonymized form (\texttt{release/})---the 154
action labels map to stable opaque codes, identifiers are hashed under a
private salt, and tool-argument payloads and free-text rationales are
stripped, with independent leak gates keeping the result clean
(\texttt{verify\_release.py}); the label mapping and the salt are withheld
together with the raw export. Two public-data chains extend the pipeline
beyond the SOC corpus and re-run from a clean host with one command each:
\texttt{chain\_pub.sh} (fetch $\to$ XES parse $\to$ fold statistics
$\to$ tables/figures) and the tool-trajectory chain
(\texttt{pub\_tool\_fetch.py} $\to$ \texttt{pub\_tool\_parse.py} $\to$
\texttt{pub\_e2\_build.py} $\to$ \texttt{pub\_e2\_train.py}, with the
two analysis add-ons \texttt{pub\_markov.py} and
\texttt{pub\_coupon.py} writing \texttt{pub\_markov\_results.json} and
\texttt{pub\_coupon.json}); every
downloaded artifact is pinned by sha256 in its fetch manifest (App.~B),
and both pipelines fold through the identical rules
(\texttt{fold\_rules.py}, imported by \texttt{parse\_real\_corpus.py},
\texttt{pub\_e1.py} and \texttt{pub\_e2\_build.py};
\texttt{e2\_build.py} carries the same rules as its original
\texttt{fold\_events}), with a lineage self-test in
\texttt{pub\_e2\_build.py} asserting its decision construction agrees
with the fold kind-for-kind. The JSONL schema and fold tooling document
the rest of what a replication needs: \texttt{parse\_real\_corpus.py}
consumes any Langfuse-style observations export; the derived-outcome rule
of \S\ref{sec:setup} is a pure function of the parsed call structure
(labels, arguments and ordering), reproduced by
\texttt{label\_real\_v2.py} on the internal \texttt{episodes.jsonl}; the
derived outcomes ship precomputed
(\texttt{outcome\_labels\_v2.jsonl}) because the release strips the
argument payloads the rule reads. Models: MIMO Pro V2.6 and DS V4.1 Flash are production
services behind our gateway (\S\ref{sec:setup}); Laya 0.3.21 is public
(pypi package \texttt{laya}, checkpoint \texttt{convaiinnovations/laya}
via ModelScope), loaded from a local snapshot. Seeded fits fix
\texttt{random\_state=0} and torch seeds 0--4; the hidden-state probe and
the E4 refits train on the full 12{,}613-decision train split (plus, for
the E4 refits, the batches accepted so far; an earlier
6{,}000-row subsample of the train split was retired before the reported runs; see the
released \texttt{e2\_laya\_probe.py}) and all arms
evaluate on the full test split (3{,}875 decisions). Timing is
wall-clock batch-1 on GPU or in-vivo service latency and will vary by
host and load. The one synthetic element (E4's 12\% corruption injection)
is regenerated from its stated seed by the E4 script.

The code release and the pseudonymized derived corpus (\texttt{release/}) are available at
\url{https://github.com/geilihan/niocis}, with a per-file \texttt{MANIFEST.sha256}; the
derived corpus is released under CC~BY~4.0.

\bibliographystyle{iclr2026_conference}
{\raggedright
\bibliography{references}
\par}

\appendix
\section*{Appendix A: fold statistics of the production corpus}
The fold of \S\ref{sec:method} decomposes the 22{,}192 logged tool calls
into 15{,}704 advance (forward) hops and 6{,}488 loop events (29.2\% of
steps). The 4{,}124 retry events arrive in 2{,}164 same-label repeat
runs (a run of $k\geq2$ consecutive identical labels contributes $k-1$
retry events, so the length classes sum to 2{,}164 runs, contributing 4{,}124
retry events): 1{,}394 runs of length 2 (median length 2), a smooth decay
(354 of length 3, 147 of length 4, 102 of length 5) and a thin tail of
167 runs of lengths 6 to 20 (20 consecutive same-label steps at the
maximum). The 2{,}364 back events return to an earlier kept label at
short-but-real distances (steps between the target's previous occurrence
and the return): 681 at $\le$1, 666 at 2--3, 553 at 4--8, 464 at 9+.
Pair roles across the fold: 2{,}315 distinct pairs fire as first-visit
forward transfers, 702 as returns (481 in both roles, 221 exclusively
returns). Domain slices carry distinct closure behavior: detection closes
1{,}001/1{,}174 episodes (85\%), response 62/245 (25\%), recognition 7/54
(13\%), protection 0/139---protection-stage sessions in this window
hand off to sibling stages rather than declare completion (a real
stage-behavior property of the deployment, not a labeling artifact; the
derived labels and their validation are in \S\ref{sec:setup}). The one
synthetic element is E4's corruption model: round 3 flips 12\% of update
labels to a uniformly drawn non-target option---the draw is seeded by the
script's per-round RNG (\texttt{RandomState(100{+}round)}, i.e.\ 103 for
round 3)---a deliberate stand-in for upstream policy regressions.
The multi-chain study quoted in the E4 figure caption runs ten refit chains: five coupled seed offsets (the torch init, the corruption draw and the validation-split permutation all shifted together) run as two 5-chain arms---one under the row-level stream split (seed
offsets 0/1000/2000/3000/4000) and one under the whole-trace split
(1000--5000; four of the five offsets shared). Each chain makes four candidate updates, so the study reports 13 refusals in 40 attempts over 6 of the 10 chains, with held-out accuracy averaging 30.3\% $\to$ 31.0\% ($\pm 0.5$ points) and an unpaired 0.02-point gap between the arms (0.09 points on the four shared offsets). Three of the ten chains dip below their previous
committed score mid-study (two of those below their base fit) before
recovering; every chain ends at or above its base fit. Results are in \texttt{e4\_rsi\_multi.json} (Reproducibility).

\section*{Appendix B: services and statistical details}
The opponent (MIMO Pro V2.6) and the judge (DS V4.1 Flash) are served through our production gateway (vLLM
backend; both are reasoning models that emit reasoning traces; decoding at
temperature 0 with a 2{,}048-token completion budget). In-vivo call
latencies appear in \S\ref{sec:results}; smoke probes on tiny next-hop
prompts run at p50 0.5\,s (MIMO Pro V2.6) and 0.45\,s (DS V4.1 Flash), growing with prompt
and reasoning length (episode-level judge calls in
\S\ref{sec:setup} ran at p50 2.0\,s, p90 5.6\,s). The typed encoder's
backbone is Laya 0.3.21 with the \texttt{convaiinnovations/laya} checkpoint
(ModernBERT-large encoder; marker-position decision head; an
accept/escalate head with cost ratio escalate 0.5 versus wrong 3.0---matching
the gap metric's $c_3{=}3$ wrong-hop weight, which does not price
escalations). Its shipped per-bucket
decision temperatures include one out-of-range entry
(\texttt{choice:11+} $= 0.1$, outside the library's stated $[0.5, 5]$
range), so reported confidences use a scalar temperature refit by
validation NLL for the zero-shot arm (all trained arms
report raw masked-renorm confidence). That refit is unbounded: its
optimum sits at whatever ceiling the grid imposes---$T{=}3.0$ on the
published eight-point grid and still $T{=}1000$ on a 60-point grid spanning
$0.3$--$1000$---so the arm's ECE-15 is grid-dependent ($0.031$ against
$0.010$) while its argmax and risk--coverage curve are unchanged. The hidden-state probe attaches a linear head to the
encoder's designated readout position---the pooled position the
checkpoint's own action head reads---and trains with CE versus CE-plus-Brier
(Adam~\citealp{kingma2015adam}, 90-epoch fits for both objectives, batch 512---the released
\texttt{e2\_laya\_probe.py} defaults) on the full
12{,}613-decision train
split, with 5 seeds (0--4) for error bars (the feature-MLP readouts use 60-epoch fits
over these seeds; the E4 and slice refits reuse the probe's frozen encoder features
with 60-epoch fits).
Trained readout time budgets: feature arms fit in seconds on CPU; the
probe's one-time encoder extraction covers the full train/test splits
(12{,}613 + 3{,}875 decisions) in 512-text chunks on GPU (throughput is
host-specific). Model
latency is wall-clock batch-1 on one NVIDIA H20 (in the
\texttt{nvcr.io/nvidia/pytorch:26.03-py3} container): typed-choice head
predict 19.5\,ms p50 (encoder included) over 17 declared options, the
measured prompt's count (50 warm trials; held-out states declare 40.2
options on average), bare encoder forward 10.0\,ms p50 (the probe's marginal
per-state cost), and, in vivo and on CPU, 0.30\,s p50 per state for the
full Laya typed-choice pipeline (the zero-shot arm)
(\texttt{gpu\_latency.json}, \texttt{e2\_laya\_results.json}); service
latencies vary under production load.

\begin{figure}[!htbp]
\centering
\includegraphics[width=\linewidth]{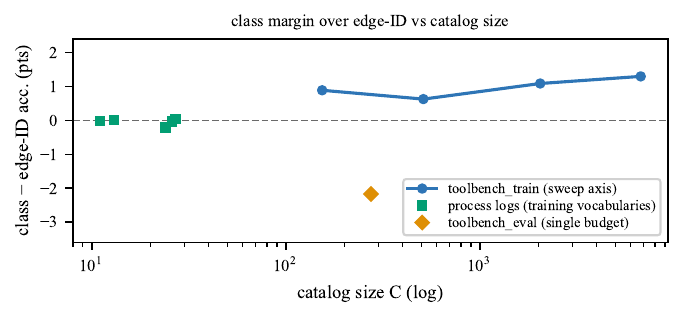}
\caption{Class-space margin over the edge-ID readout versus catalog
budget (top-$C$ train-frequency actions; masked accuracy, 3 of the 5 protocol seeds: both arms are refit at
every budget point under \texttt{pub\_sweep.py}'s seeds-0/1/2
convention; the main-text reading and exception accounting are in \S\ref{sec:results} and App.~B). The
7{,}318-name ToolBench catalog~\citep{qin2023toolllm} provides the only multi-budget axis: the
class vocabulary's margin is non-monotone but ends wider (never
crossing zero in the plotted 3-seed means; per-arm seed standard
deviations stay under 0.7 points; net figures in text)---the
opposite of the edge-pays-at-scale prediction.
Squares mark the five process logs at their 11--27-name training vocabularies
(noise-level ties); the single diamond is the evaluation subset at its
275-name training catalog, the only plotted contrary slice (2.2 points to
edge IDs over 92 of its 161 held-out hops, scored in-catalog---2.4 points
on the fixed-row paired rerun). API-Bank is a second small-catalog reversal (1.7 points to edge IDs under the
paired convention of the fixed-row rerun---one whose 335-row training fold sits
below that sweep's own floor and whose paired gap is one-seed-driven, per-seed
differences 0.0/0.0/5.2 points---and 3.5 points on unrounded means in the CE-arm pair of the main
API-Bank runs, \S\ref{sec:results}).}
\label{fig:sweep}
\end{figure}
\paragraph{Public corpus provenance (E5).}
Six event logs, fetched 2026-09-29 from their canonical hosts (some via pinned
mirrors) and pinned by
sha256 (\texttt{pub\_fetch\_manifest.json}): receipt, roadtraffic and
helpdesk (\texttt{pm4py-core} test fixtures~\citealp{berti2023pm4py} via the gh-proxy mirror),
BPIC'12 (pm4py \texttt{bpic2012.xes.gz}), BPIC'17 and the Statechart
Workbench software session (the 4TU file service; session log of \citet{leemans2018statechart}). The tool-trajectory arm
uses API-Bank~\citep{li2023apibank} (263 scenario files from
\texttt{AlibabaResearch/DAMO-ConvAI}) and
ToolBench~\citep{qin2023toolllm} \texttt{toolllama\_G123\_dfs} chains
(Adorg/ToolBench mirror on
hf-mirror.com; 300 eval and 26{,}999 train trajectories), sha256-pinned in
\texttt{tool\_manifest.json}. Two candidate corpora were deliberately dropped
rather than approximated---Zenodo discovery only surfaced a course homework
scaffold for the Sepsis cases log of \citet{mannhardt2016sepsis} and a paper's \texttt{inputs.zip} bundle for BPIC'11 (2011
BPI Challenge~\citealp{vandongen2011hospital})---and the xLAM 60k function-calling corpus~\citep{liu2024apigen} stays behind a
HuggingFace LFS gate that denies anonymous fetches; no domestic
mirror hosts that corpus. All three are simply absent: a
missing row beats uncertain lineage. A builder-correction ledger accompanies the pipeline:
URLs are verified against host APIs before hardcoding; extension-less 4TU
links are typed by content magic; Content-Length and full gzip/zip
integrity checks caught two silently truncated downloads (17.6\,MB of
29.7\,MB and 6.8\,MB of 147\,MB) before any number was computed from
them; per-URL watchdogs guard against resolver hangs; and an XES trace-ID
fallback covers software logs (which label traces
\texttt{concept:instance}, not \texttt{concept:name}). Every reported
number travels through verified, integrity-checked local copies.

\paragraph{Transition-table baseline.} Contexts over the traced node
flow (orders 0--3) are counted on the training fold of each of the nine
decision streams (SOC; API-Bank, BPIC'12, BPIC'17, helpdesk, receipt,
roadtraffic and the two ToolBench streams); each order interpolates down
to the training marginal with a cascading
weight $\lambda=w_{\mathrm{rem}}\, n_{\mathrm{ctx}}/(n_{\mathrm{ctx}}{+}3)$
($w_{\mathrm{rem}}$: weight left unspent by higher orders, cascading
3$\to$2$\to$1, residue to the training marginal) and the argmax is
restricted to the declared options (the deployed masked-renorm gate).
Two scorings bracket each stream. The verbatim majority arm (raw
per-node argmax)---scored raw and option-masked (a hit only when the
argmax lands in the options)---recovers the \S\ref{sec:setup} majority arm to the digit and
matches every other stream's identically-built look-up-table (LUT) arm
exactly; the interpolated tables (an interpolated n-gram with backoff
semantics---the backoff Markov) gate before scoring. For the
majority arm, test accuracy equals its option-masked accuracy on all
nine streams (\texttt{pub\_markov\_results.json}): option-gating never
removes a correct majority prediction. The label catalog is
treated as known (the class index covers every corpus label, held-out
names included, so no action target is out of vocabulary). The public
row encoding would bucket one at \texttt{<stop>}, and the edge-ID arms
map training-fold-unseen edge targets to \texttt{<unk\_edge>}; option
sets and edge identifiers come from the training fold only.

\paragraph{Occupancy-null computation.}
{\emergencystretch=3em\relax\raggedright The dashed occupancy null of
\mbox{\Cref{fig:pubsat}a} evaluates $\mathbb{E}[D(m)]$ at the same subsample
sizes from per-label case supports alone---no fitted parameters
(\texttt{pub\_coupon.json}); its p50-versus-expectation comparison carries two residual components
by design: the 20-draw median sampling noise of the panels and any
mean-versus-median mismatch ($\mathbb{E}[D(m)]$ need not equal the p50).\par}

\paragraph{Catalog-size sweep (the edge-pays-at-scale hypothesis,
\S\ref{sec:limitations}).}
\texttt{pub\_sweep.py} holds the fold, features and training procedure fixed
while shrinking the option vocabulary to the top-$C$ actions by train
frequency, scoring only in-catalog decisions and rebuilding the option
graph and edge vocabulary inside the catalog. Sweep budgets are bounded
by each corpus's training vocabulary: the five process logs contribute
single points at their training vocabularies (11 to 27 names,
$|\Delta| \leq 0.21$ points, noise-level ties) and ToolBench contributes the axis:
the class-space readout beats the edge-ID readout at every budget ($C{=}154$: 61.5\%
versus 60.6\%; $C{=}512$: 54.3 versus 53.7; $C{=}2{,}048$: 47.6 versus
46.5; $C{=}6{,}750$: 43.2 versus 41.9). Training floors: the 335-row
API-Bank training fold and the 154-name slice of the evaluation subset fall below
the sweep's 500-train/50-test guard, so those rows are skipped.

\paragraph{The induced graph read as a discovered model.}
For completeness we report the four standard discovery dimensions for the
induced graph ($\S\ref{sec:related}$), with conventions stated because they are
convention-dependent: \emph{fitness} is the share of traces whose every
consecutive kept pair is a model edge, with a token-replay flavour
($1-\text{unmatched pairs}/\text{pairs}$); \emph{precision} is the
escaping-edge form (ETConformance style: $1-\sum\text{escaping}/\sum\text{enabled}$
over trace prefixes, where escaping counts enabled outgoing edges not taken);
\emph{generalization} is the counting approximation
($1-\text{mean}\,1/\sqrt{\text{visits}}$ over model elements); \emph{simplicity}
is reported as counts (143 nodes with outgoing edges, 2{,}517 edges---1{,}944
forward, 573 recovery---mean out-degree 17.6). The graph is induced from the
\emph{train} fold, so in-sample readings are degenerate by construction
(trace- and token-level fitness $1.000$); on held-out folds token-level fitness
is $0.902$ (val) and $0.913$ (test, unmatched pairs $307/3{,}530$), trace-level
fitness $0.621$ and $0.591$, escaping-edge precision $0.022$ and $0.021$, and
generalization is $0.625$ at node level against $0.267$ at edge level. The
permissive precision is structural: each state enables its entire declared
option set ($\bar{|A(v)|}=40.2$), so precision tracks $1/\bar{|A(v)|}$. No
inductive-miner comparison is attempted, and these numbers are not offered as a
process-discovery contribution.

\end{document}